\documentclass[aps,prd,preprint,showpacs,showkeys,nofootinbib,superscriptaddress]{revtex4-1}

\usepackage{amsmath}
\usepackage{amssymb}
\usepackage{dcolumn}
\usepackage{bm}
\usepackage{feynmp}                                     
\usepackage{graphicx}
\usepackage[breaklinks=true]{hyperref}
\usepackage{breakurl}           
\usepackage{subcaption}
\usepackage{color}

\begin{document}

\title{Testing the Scalar Sector of the Twin Higgs Model at Colliders}

\author{Zackaria Chacko}
\affiliation{Maryland Center for Fundamental Physics, Department of Physics,\\ University of Maryland, College Park, MD 20742-4111 USA}
\affiliation{Theoretical Physics Department, Fermilab, P.O. Box 500, \\
Batavia, IL 60510, USA} 
\author{Can Kilic}
\affiliation{Theory Group, Department of Physics and Texas Cosmology Center\\ University of Texas at Austin,
Austin, TX 78712, USA}
\author{Saereh Najjari}
\affiliation{Faculty of Physics, University of Warsaw, Pasteura 5, 02-093 Warsaw, Poland}
\author{Christopher B. Verhaaren}
\affiliation{Center for Quantum Mathematics and Physics (QMAP), Department of Physics,\\ University of California, Davis, CA, 95616-5270 USA.}

\date{\today}
\begin{abstract}

We consider Mirror Twin Higgs models in which the breaking of the global 
symmetry is realized linearly. In this scenario, the radial mode in the 
Higgs potential is present in the spectrum, and constitutes a second 
portal between the twin and SM sectors. We show that a study of the 
properties of this particle at colliders, when combined with precision 
measurements of the light Higgs, can be used to overdetermine the form 
of the scalar potential, thereby confirming that it possesses an 
enhanced global symmetry as dictated by the Twin Higgs mechanism. We 
find that, although the reach of the LHC for this state is limited, 
future linear colliders will be able to explore a significant part of 
the preferred parameter space, allowing the possibility of directly 
testing the Twin Higgs framework.

\end{abstract}

\preprint{UTTG-10-17}



\maketitle

\section{Introduction\label{s.intro}}

The discovery of the Higgs boson at 
CERN~\cite{Aad:2012tfa,Chatrchyan:2012ufa}, while completing the 
standard model (SM), has also brought the hierarchy problem, the 
question of the radiative stability of the Higgs mass, into sharper 
focus. Symmetry-based solutions to the hierarchy problem~\cite{Fayet:1977yc,Dimopoulos:1981zb,ArkaniHamed:2001nc} require new 
particles with masses at or below the TeV scale that have sizable 
couplings to the Higgs. Searches for these particles continue to be a 
major focus of the Large Hadron Collider (LHC) program. These analyses 
have become increasingly powerful and sophisticated in an effort to 
explore all simple realizations of the known cancellation mechanisms. So 
far, these searches, while imposing stiff constraints on solutions to 
the hierarchy problem, have provided no hints as to its resolution.

One explanation of these null results is that while the Higgs mass is 
indeed protected by a symmetry, the new particles associated with this 
symmetry are not charged under SM color. These states are then much more 
difficult to produce at a hadron collider, which complicates efforts to 
discover them. Several theories of this type have been proposed that 
stabilize the Higgs mass up to scales of order 5-10 TeV, the precision 
electroweak scale~\cite{Chacko:2005pe,Barbieri:2005ri,Chacko:2005vw,Burdman:2006tz,Cai:2008au,Poland:2008ev,Batell:2015aha,Serra:2017poj,Csaki:2017jby}. The best known example of this class of models is the 
Mirror Twin Higgs (MTH)~\cite{Chacko:2005pe}, in which the symmetry 
partners are neutral, not just under SM color, but under all the SM 
gauge groups.

In the MTH framework, the particle content of the SM is extended to 
include a mirror (``twin") copy of all the fields in the SM. A discrete 
$Z_2$ twin symmetry relates the particles and interactions in the SM and 
mirror sectors. The Higgs sector respects a larger global symmetry 
which, in the simplest incarnation of the model, is taken to be 
SU(4)$\times$U(1). This global symmetry, like the discrete symmetry, is only approximate.
The electroweak gauge symmetries of the 
SM and twin sectors are embedded inside the global symmetry. The fields 
that constitute the SM Higgs doublet are among the 
pseudo-Nambu-Goldstone bosons (pNGBs) associated with the spontaneous 
breaking of the global SU(4)$\times$U(1) symmetry down to 
SU(3)$\times$U(1). Their mass is protected against one loop radiative 
corrections by the combination of the nonlinearly realized global 
symmetry and the discrete twin symmetry.

Since the original proposal, the MTH scenario has been further 
developed. Ultraviolet completions based on 
supersymmetry~\cite{Falkowski:2006qq,chang:2006ra,Craig:2013fga,Katz:2016wtw,Badziak:2017syq,Badziak:2017kjk} 
and Higgs 
compositeness~\cite{Geller:2014kta,Barbieri:2015lqa,Low:2015nqa} that 
can raise the cutoff to the Planck scale have been proposed, and their 
collider implications studied~\cite{Cheng:2015buv,Cheng:2016uqk}. 
Composite Twin Higgs models 
have been shown to be consistent with precision electroweak constraints~\cite{Contino:2017moj} 
and flavor bounds~\cite{Csaki:2015gfd}. Various possibilities for the 
breaking of the discrete 
$Z_2$ symmetry and their effects on  
tuning 
have been studied~\cite{Beauchesne:2015lva,Harnik:2016koz,Yu:2016bku,Yu:2016swa,Yu:2016cdr}. 
More general global symmetry breaking patterns have also been 
investigated~\cite{Craig:2014aea,Craig:2014roa,Craig:2016kue,Thrasher:2017rpa}.

The cosmology of the MTH model is rather problematic. The twin sector is 
in thermal equilibrium with the SM in the early universe up to 
temperatures of order a few GeV~\cite{Barbieri:2005ri}. At lower 
temperatures the two sectors decouple, but the twin photon and twin 
neutrinos survive as thermal relics. These states contribute 
significantly to the total energy density in radiation, 
which conflicts with the bounds on dark radiation from the cosmic microwave 
background and 
Big Bang nucleosynthesis. This problem can be solved 
if the model is extended to realize a new contribution to the energy 
density of the SM sector after the two sectors have 
decoupled~\cite{Chacko:2016hvu,Craig:2016lyx}. In general, this does not 
require additional breaking of the discrete 
$Z_2$ symmetry. 
Alternatively, 
the problem can be solved by introducing hard breaking of 
the
$Z_2$ 
into the twin sector Yukawa couplings, 
thereby altering the spectrum of mirror 
states~\cite{Farina:2015uea,Barbieri:2016zxn,Csaki:2017spo,Barbieri:2017opf}. 
Once these cosmological bounds are satisfied puzzles like the nature of 
dark matter~\cite{Farina:2015uea} or the baryon 
asymmetry~\cite{Farina:2016ndq} can be addressed.

Recently an alternative class of Twin Higgs models, known as Fraternal 
Twin Higgs (FTH) models, has been proposed, in which the twin sector is 
more minimal than in the MTH, consisting of
only those states that are 
required to address the hierarchy problem~\cite{Craig:2015pha}. 
Specifically, the spectrum of light twin sector states 
includes only the 
third generation 
fermions, the electroweak gauge bosons, and the twin 
gluon. This framework naturally solves the cosmological problems of the 
MTH construction. 
It also leads to exotic collider signals since 
the lightest twin particles, the mirror glueballs, decay back to SM 
states, but with long lifetimes. Mirror glueballs can be produced in 
Higgs decays and will then decay far from the original interaction 
point, resulting in displaced vertices. The striking nature of these 
signals will allow the LHC to probe most of the preferred parameter 
space~\cite{Curtin:2015fna,Csaki:2015fba}. The proposed MATHUSLA 
detector~\cite{Chou:2016lxi,Curtin:2017izq} is also expected to be 
sensitive to the displaced decays arising from this class of models.  
The FTH also contains several promising dark matter 
candidates~\cite{Craig:2015xla,Garcia:2015loa,Garcia:2015toa,Freytsis:2016dgf} 
and has been put forward as a possible explanation of certain observed 
anomalies in large and small scale structure~\cite{Prilepina:2016rlq}.

The only communication between the visible and twin sectors that is 
required by the Twin Higgs framework is through the Higgs portal. After 
electroweak symmetry breaking the Higgs fields of the two sectors mix. 
The lighter mass eigenstate is identified with the 125 GeV Higgs 
particle. As a consequence of the mixing it has suppressed couplings to 
SM fields, resulting in a production cross section that is smaller than 
the SM prediction. This mixing also results in a contribution to the 
Higgs width from decays into invisible twin sector states. 
Unfortunately, while these signals are robust predictions of the MTH 
framework, they are not unique to it. They are expected to arise in any 
model in which the SM communicates with a light hidden sector through 
the Higgs portal.

If, however, the $Z_2$ symmetry is only softly broken, so that the 
Yukawa couplings in the two sectors are equal, the suppression in the 
Higgs production cross section and the Higgs invisible width are both 
determined by the mixing angle, leading to a prediction that can be 
tested by experiment~\cite{Burdman:2014zta}. This prediction does not 
apply to theories that exhibit hard breaking of $Z_2$, such as the FTH, 
or MTH models in which the Yukawa couplings in the two sectors are 
different. The prediction can be understood as a consequence of the 
mirror nature of the model. Since it does not depend on the enhanced 
global symmetry of the Higgs sector, this prediction is not specific to 
the MTH construction, but applies more generally to any mirror model~\cite{Foot:1991bp,Foot:1991py} in 
which the discrete $Z_2$ symmetry is only softly broken, so that the
Yukawa couplings in the two sectors are equal.

If the breaking of the global symmetry is realized linearly, the radial
mode in the Higgs potential is present in the spectrum and constitutes
a second portal between the twin and SM sectors. We refer to this state
as the twin sector Higgs. As we now explain, a study of the properties
of this particle at colliders, when combined with precision measurements
of the light Higgs, can be used to overdetermine the form of the scalar
potential, thereby confirming that it possesses an enhanced global
symmetry as dictated by the Twin Higgs mechanism.

In the case when the discrete $Z_2$ symmetry is only softly broken, the 
Higgs potential of the MTH model takes the form{\footnote{We employ the 
notation of~\cite{Barbieri:2005ri}.}}
 \begin{align}
V=&-\mu^2\left(H_A^\dag H_A+H_B^\dag H_B \right)+\lambda\left(H_A^\dag H_A+H_B^\dag H_B \right)^2\nonumber\\
&+m^2\left( H_A^\dag H_A-H_B^\dag H_B\right)+\delta\left[\left( H_A^\dag H_A\right)^2+\left(H_B^\dag H_B \right)^2 \right].
\label{e.higgspot}
 \end{align}
We distinguish the SM sector fields with the subscript $A$ and the twin 
sector fields with $B$. The terms in the top line of 
Eq.~\eqref{e.higgspot} respect both the global SU(4)$\times$U(1) 
symmetry and the discrete $Z_2$ twin symmetry $A\leftrightarrow B$. The 
$m^2$ term explicitly breaks both the discrete and global symmetries, 
but only softly, and can naturally be smaller than $\mu^2$. The quartic 
term $\delta$ respects the $Z_2$ twin symmetry, but violates the global 
symmetry. In order to realize the light Higgs as a pNGB and thereby obtain a significant reduction in 
fine-tuning 
relative to the SM the parameter 
$\delta$ that violates the global symmetry 
must be much smaller than $\lambda$, which is invariant 
under SU(4)$\times$U(1). Similarly, $m^2$ must be much smaller than 
$\mu^2$.

The parameters in the 
Higgs potential must reproduce the mass of the 
light Higgs and the electroweak vacuum expectation value (VEV). This fixes two combinations of the 
four parameters. Two additional measurements are then required to fully 
determine the 
potential. At a lepton collider the production cross 
section and invisible width of the light Higgs can be determined to a 
precision of order one part in a 
hundred~\cite{Fujii:2015jha,Abramowicz:2016zbo}. This covers the entire 
range of interest for the MTH and fixes a third combination of the 
parameters. Finally, the discovery of the twin sector Higgs particle at 
a given mass would pin down all four parameters in the Higgs potential. 
Once the potential has been specified, in the absence of further $Z_2$ 
violation, the production cross section, width and branching ratios of 
the twin sector Higgs are all robustly predicted. Therefore, a 
measurement of the rate to any SM final state overdetermines the system, 
and constitutes a powerful consistency check on the form of the 
potential. These predictions remain true to a good approximation even in 
the presence of hard breaking of the $Z_2$ symmetry by the twin sector 
Yukawa couplings, provided that this breaking is not large enough to 
significantly alter the total width of the twin sector Higgs.

In the MTH framework, the breaking of the approximate global symmetry of 
the Higgs potential results in seven pNGBs. These 
include, in addition to the light Higgs, the longitudinal components of 
the $W^{\pm}$ and $Z$ bosons of both the SM and twin sector. It follows 
that in the limit that the global symmetry is exact, the couplings of 
the twin sector Higgs particle to all these seven states are the same, and 
are set by the SU(4)$\times$U(1) invariant quartic term in the Higgs potential. In particular, the 
couplings of this state to the SM Higgs, $W^{\pm}$, and $Z$ are not 
suppressed by the mixing angle. In the limit that the twin sector Higgs 
particle is heavy, corresponding to the quartic term being large, its dominant decay modes are to these seven 
pNGBs. Furthermore, in the limit that the masses of 
the final state particles can be neglected, the branching ratio of the 
twin sector Higgs into each of these final states is the same. It 
follows that $WW$, $ZZ$ and di-Higgs are promising channels in which to 
search for the twin sector Higgs.
 
In this paper we study the prospects for the LHC and future colliders to 
discover the twin sector Higgs and determine the form of Higgs 
potential, thereby confirming the MTH framework. Discovery of the twin 
sector Higgs scalar at the LHC has been discussed 
previously~\cite{Craig:2015pha,Buttazzo:2015bka,Craig:2016kue,Katz:2016wtw,Ahmed:2017psb}, 
but without the emphasis on 
determining the structure of the potential. 
We find that at the LHC, much of the range of parameter space in which 
the twin sector Higgs can be discovered is already disfavored by 
existing measurements of the couplings of the light Higgs. Only a 
restricted set of parameters leads to Higgs coupling deviations and a 
twin sector Higgs signal that can both be measured at the LHC. We find 
that the high energy stages of linear colliders such as the 
international linear collider (ILC)~\cite{Behnke:2013xla} or compact 
linear collider (CLIC)~\cite{CLIC:2016zwp} are expected to have much 
greater reach for the twin sector Higgs. These colliders are also 
projected to measure the invisible width of the light Higgs to percent 
level precision~\cite{Fujii:2015jha,Abramowicz:2016zbo}. Combining the 
measurements of the twin sector Higgs with precision studies of the 
couplings of the light Higgs results in much greater ability to confirm 
the MTH construction.

The outline of this paper is as follows. In Sec.~\ref{s.nonlintwin} we 
describe the scalar sector of the MTH in detail, and develop the 
notation we use in the rest of the paper. In Sec.~\ref{s.LHC}, we 
determine the reach of the LHC for the twin sector Higgs. In 
Sec.~\ref{s.LinCol}, we study the potential for the ILC and CLIC to 
discover the twin sector Higgs, and determine its couplings. We conclude 
in Sec. V.

\section{The Scalar Sector of the Mirror Twin Higgs\label{s.nonlintwin}}

This section outlines the dynamics of the scalar sector in the MTH 
framework. We begin by analyzing the Higgs potential shown in 
Eq.~\eqref{e.higgspot}. It is convenient to employ an exponential 
parametrization of the scalar degrees of freedom. Accordingly, we define 
an object $H$ which transforms linearly under SU(4)$\times$U(1),
 \begin{equation}
H=\left(\begin{array}{c}
H_A\\
H_B
\end{array} \right) =
\exp\left(\frac{i}{f}\Pi \right)\left(\begin{array}{c}
0\\
0\\
0\\
\displaystyle f+\frac{\sigma}{\sqrt{2}}
\end{array} \right) \; .
 \end{equation}
 Here $f$ is the symmetry breaking VEV, and $\Pi$ is given, in unitary 
gauge where all the $B$ sector pNGBs have been eaten by the corresponding 
vector bosons, by
 \begin{equation}
\Pi=\left(\begin{array}{ccc|c}
0&0&0&ih_1\\
0&0&0&ih_2\\
0&0&0&0\\ \hline
-ih_1^{\ast}&-ih_2^{\ast}&0&0
\end{array} \right).
 \end{equation}
Expanding the exponential we obtain
 \begin{equation}
H=\left(\begin{array}{c}
\displaystyle\frac{\bm{h}}{\sqrt{\bm{h}^{\dag}\bm{h}}}\left( f+\frac{\sigma}{\sqrt{2}}\right)\sin\left( \frac{\sqrt{\bm{h}^{\dag}\bm{h}}}{f} \right)\\
0\\
\displaystyle \left(f+\frac{\sigma}{\sqrt{2}}\right)\cos\left( \frac{\sqrt{\bm{h}^{\dag}\bm{h}}}{f} \right)
\end{array} \right) \; ,
\end{equation}
where $\bm{h} = (h_1, h_2)^{\text{T}}$ is the SM Higgs doublet. Proceeding to unitary gauge in the SM sector with $h_1=0$ and $h_2=(v+h)/\sqrt{2}$ leads to
 \begin{align}
\displaystyle H_A&= \left( \begin{array}{c}
0\\
\left(f+\frac{\sigma}{\sqrt{2}} \right)\sin\left( \frac{v+h}{\sqrt{2}f} \right)
\end{array}\right), \\
 H_B&= \left(\begin{array}{c}
0\\
\left(f+\frac{\sigma}{\sqrt{2}} \right)\cos\left( \frac{v+h}{\sqrt{2}f}  \right)
\end{array} \right),
\end{align}
and allows us to write the potential as
 \begin{align}
V=&f^2\left(1+\frac{\sigma}{\sqrt{2}f}\right)^2\left[ -\mu^2-m^2\cos\left( \frac{\sqrt{2}(v+h)}{f}\right)\right]\nonumber\\&+f^4\left(1+\frac{\sigma}{\sqrt{2}f}\right)^4\left[\lambda+\delta -\frac{\delta}{2}\sin^2\left( \frac{\sqrt{2}(v+h)}{f}\right) \right]. 
\label{e.nonlinpot}
 \end{align}
 It is convenient to define the angular variable $\vartheta \equiv 
v/(\sqrt{2}f)$. In terms of $\vartheta$ and $f$, the VEVs in the visible 
and twin sectors are given by
 \begin{equation}
v_\text{EW}\equiv\sqrt{2}f\sin\vartheta , \ \ \ \ v_B\equiv\sqrt{2}f\cos\vartheta \;.
 \end{equation}
 The equations of motion for $\sigma$ and $h$ take the form 
 \begin{align}
\mu^2+m^2\cos(2\vartheta)&=f^2\left[2(\lambda+\delta)-\delta\sin^2(2\vartheta) \right],\\
m^2&=\delta f^2\cos(2\vartheta).\label{e.mCondition}
 \end{align}
 We see from the second equation that in the $Z_2$ symmetric limit $(m^2 
= 0)$ the mixing is maximal, with $v_{\text{EW}}=v_B$ and 
$\theta=\pi/4$. This would mean that the observed 125 GeV Higgs boson 
couples just as strongly to the twin sector as to the SM, which  
conflicts with current data. Therefore we need $m^2 > 0$, which 
corresponds to $\vartheta < \pi/4$, to obtain realistic phenomenology. We have assumed here that $\delta>0$, which is required for stable vacuum~\cite{Barbieri:2005ri}.
 Combining the equations of motion we obtain 
 \begin{equation}
\mu^2=f^2\left(2\lambda+\delta \right).
 \label{e.muCondition}
\end{equation}

The mass eigenstates $h_{-}$ and $h_{+}$ are linear combinations of $h$
and $\sigma$,  
\begin{equation}
\left(\begin{array}{c}
h_{-}\\
h_{+}
\end{array} \right)=\left( \begin{array}{cc}
\cos\theta & \sin\theta\\
-\sin\theta & \cos\theta
\end{array}\right)
\left(\begin{array}{c}
h\\
\sigma
\end{array} \right).
 \end{equation}
 The mixing angle $\theta$ is given by
 \begin{equation}
\begin{array}{cc}
\displaystyle \sin(2\theta)=\frac{2f^2\delta\sin(4\vartheta)}{m_{+}^2-m_{-}^2}, & \displaystyle \cos(2\theta)=\frac{4f^2\left[\lambda+\delta\cos^2(2\vartheta)\right]}{m_{+}^2-m_{-}^2}.
\end{array}
 \end{equation}
 The mass eigenvalues $m_{+}$ and $m_{-}$ are given by 
 \begin{equation}
m_\pm^2=2f^2\left[ \lambda+\delta\pm\sqrt{\lambda^2+\delta(2\lambda+\delta)\cos^2(2\vartheta)}\right] .
\label{masseigenvalues}
 \end{equation}
 We can express $\lambda^2$ in terms of $m_+$, $m_-$, and $\vartheta$,  
 \begin{equation}
\lambda^2=\frac{1}{16f^4}\left[(m_{+}^2-m_{-}^2)^2-4\cot^2(2\vartheta)m_{+}^2m_{-}^2 \right].
 \end{equation}
 Note that in order to keep $\lambda^2\geq 0$ we must have
 \begin{equation}
\frac{m_{+}}{m_{-}}\geq |\cot(2\vartheta)|+|\csc(2\vartheta)|.
 \end{equation}
 Since $\vartheta<\pi/4$ we can drop the absolute value symbols to obtain 
 \begin{equation}
\frac{m_{+}}{m_{-}}\geq \cot\vartheta =\frac{v_B}{v_\text{EW}} 
= \frac{m_T}{m_t} \;. 
 \end{equation}
 Here $m_t$ represents the mass of the top quark and $m_T$ the mass of
the its twin counterpart, the top partner. This inequality places a lower 
bound on the mass of the twin sector Higgs relative to the mass of the
top partner.

\subsection{Higgs Couplings in the Twin Higgs Framework\label{ss.gauge}}

The couplings of the Higgs fields to the $W$ and $Z$ gauge bosons arise 
from the kinetic terms,
 \begin{equation}
\left|\left(\partial_\mu+igW^{(j)}_{\mu,A}+\frac{i}{2}g'B_{\mu,A}\right)H_A \right|^2\;+\;\left( A\to B\right) \;.
\label{WZcouplings}
 \end{equation}
 Here the $W^{(1,2,3)}_\mu$ are the three gauge bosons of SU(2$)_L$ and 
$B_\mu$ that of hypercharge. The resulting $W$ boson masses are given by
 \begin{equation}
\begin{array}{cc}
\displaystyle M^2_{W_A}=\frac{f^2g^2}{2}\sin^2\vartheta=\frac{g^2v_\text{EW}^2}{4}, & \displaystyle M^2_{W_B}=\frac{f^2g^2}{2}\cos^2\vartheta=\frac{g^2v_B^2}{4},
\end{array}
 \end{equation}
 The $Z$ boson masses are related to these by the usual factor of 
$\cos\theta_W$. In unitary gauge, the couplings of the light Higgs to 
the $W$ and $Z$ bosons that result from Eq.~\eqref{WZcouplings} take the 
form (see the Appendix for details)
 \begin{align}
2\frac{h_{-}}{v_\text{EW}}\cos(\vartheta-\theta)&\left[M^2_{W_A}W_{A\mu}^{+}W_A^{\mu-}+\frac12M^2_{Z_A}Z_{A\mu}Z_A^{\mu} \right] \nonumber\\
-2\frac{h_{-}}{v_B}\sin(\vartheta-\theta) &\left[M^2_{W_B}W_{B\mu}^{+}W_B^{\mu-}+\frac12M^2_{Z_B}Z_{B\mu}Z_B^{\mu} \right] \;.
 \end{align} 
 The corresponding expression for the heavy Higgs is given by
\begin{align}
2\frac{h_{+}}{v_\text{EW}}\sin(\vartheta-\theta)&\left[M^2_{W_A}W_{A\mu}^{+}W_A^{\mu-}+\frac12M^2_{Z_A}Z_{A\mu}Z_A^{\mu} \right] \nonumber\\
+2\frac{h_{+}}{v_B}\cos(\vartheta-\theta) &\left[M^2_{W_B}W_{B\mu}^{+}W_B^{\mu-}+\frac12M^2_{Z_B}Z_{B\mu}Z_B^{\mu} \right] \;.
 \end{align} 

The couplings of the SM and twin sector fermions to the Higgs fields 
emerge from the Yukawa interactions,
 \begin{equation}
-Y^{ij}_u \overline{Q}_{A\alpha}^i\epsilon^{\alpha\beta}H^\dag_{A\alpha}u_A^j-Y^{ij}_d\overline{Q}_{A\alpha}^i H^\alpha_Ad_A^j-Y^{ij}_\ell \overline{L}_{A\alpha}^i H^\alpha_A e^j_A\;+\;\left( A\to B\right) \;.
\label{Fcouplings}
 \end{equation}
 Here $i$ and $j$ represent flavor indices while $\alpha$ and $\beta$ 
are SU(2) indices. This results in the fermions acquiring masses,
 \begin{equation}
\begin{array}{cc}
 m_{f_A}=Y_fv_\text{EW}/\sqrt{2},&  m_{f_B}=Y_fv_B/\sqrt{2}.
\end{array}
 \end{equation}
 The couplings of the light Higgs to the fermions $f$ that result from 
Eq.~\eqref{Fcouplings} take the form (for details see the Appendix),
 \begin{equation}
-h_{-}\left[\overline{f}_Af_A\frac{m_{f _A}}{v_\text{EW}}\cos(\vartheta-\theta)-\overline{f}_Bf_B\frac{m_{f_B}}{v_B}\sin(\vartheta-\theta) \right] \;.
 \end{equation}
 The corresponding couplings of the twin sector Higgs are given by
 \begin{equation}
-h_{+}\left[\overline{f}_Af_A\frac{m_{f _A}}{v_\text{EW}}\sin(\vartheta-\theta)+\overline{f}_Bf_B\frac{m_{f_B}}{v_B}\cos(\vartheta-\theta) \right].
 \end{equation}

 We see from this that, in general, the masses of the visible and twin 
sector particles are related by
 \begin{equation}
m_B^2=m_A^2\frac{v_B^2}{v_\text{EW}^2}=m_A^2\cot^2\vartheta.
 \end{equation}
 The couplings of the light Higgs $h_{-}$ to visible sector particles are related to
the corresponding couplings in the SM by
 \begin{equation}
g_{h_{-}SM}=g_{SM}\cos(\vartheta-\theta),
 \end{equation}
 while the corresponding couplings of the twin sector Higgs $h_{+}$ are given by
 \begin{equation}
g_{h_{+}SM}=g_{SM}\sin(\vartheta-\theta).
 \end{equation}
 The couplings of the Higgs fields to the twin sector are also related 
to the corresponding SM couplings. The light Higgs couples to twin 
states as it does to SM states, but with the replacement $v_{EW}\to 
v_B$, and a factor of $\sin(\vartheta-\theta)$. The couplings of $h_{+}$ 
to twin states are again those of the SM, but with the replacement 
$v_{EW}\to v_B$, and a factor of $\cos(\vartheta-\theta)$.

As detailed in the Appendix, loop induced couplings to pairs 
of photons or gluons result from a single tree level coupling between a 
Higgs and the fermion or vector in the loop, leading to the same 
$\cos(\vartheta-\theta)$ or $\sin(\vartheta-\theta)$ modification. 
However, the decays of the heavy Higgs are also a function of the mass 
of the Higgs. So, the decay widths of the twin sector Higgs to visible sector states are 
those of a SM Higgs \emph{with mass} $m_{+}$ multiplied by 
$\sin^2(\vartheta-\theta)$.

\begin{figure}[th]
\centering
\includegraphics[width=0.9\textwidth]{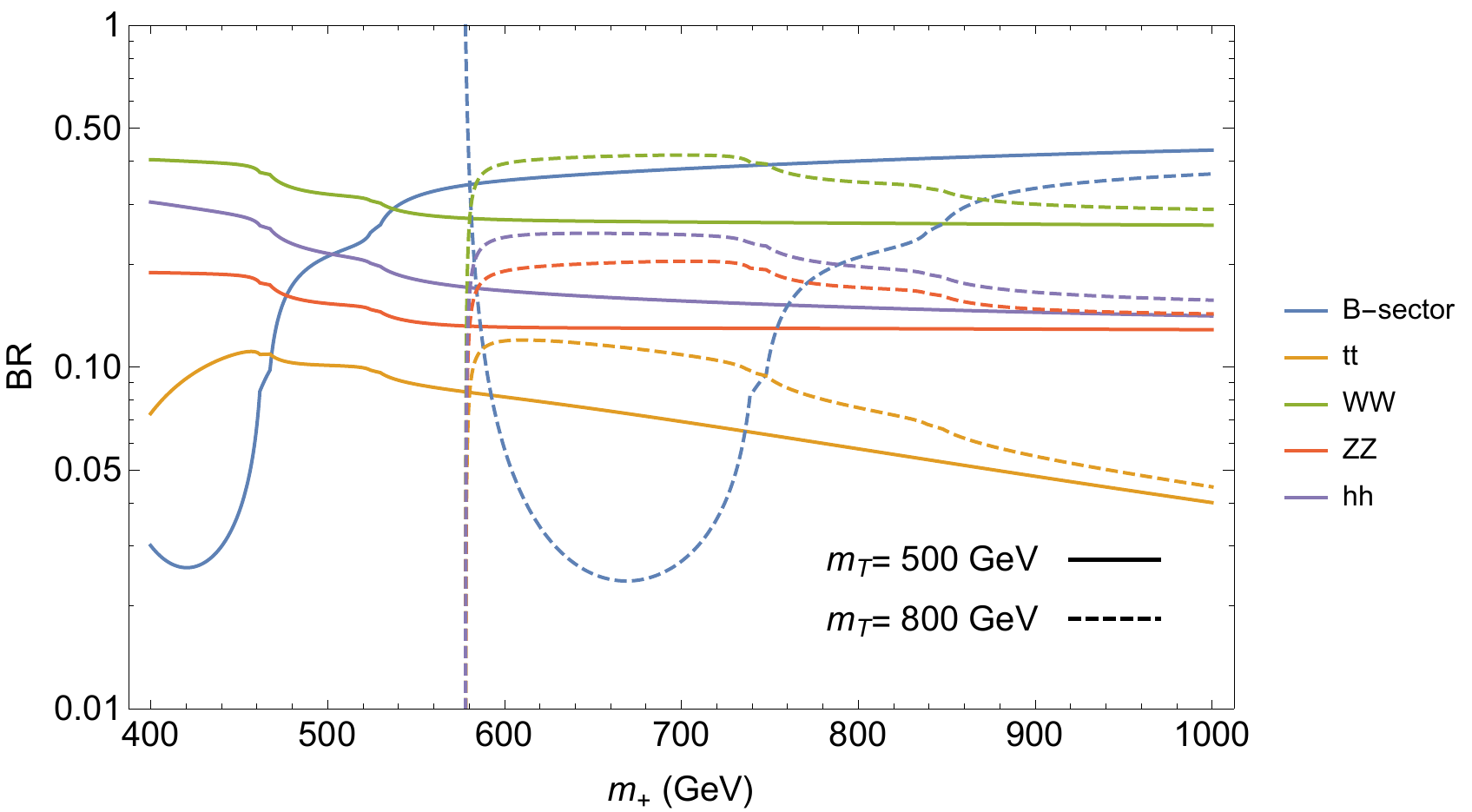}
\caption{\label{f.TwinHiggsBR}Branching ratios of the heavy twin sector Higgs scalar to various SM final states and the twin sector. }
\end{figure}

\subsection{Decays of the Twin Sector Higgs}

With the couplings of the twin sector Higgs in hand, we are now in a 
position to compute its branching ratios.
From Fig.~\ref{f.TwinHiggsBR} we see that for $m_+ < m_T$ the 
twin sector Higgs decays primarily to SM electroweak gauge bosons. Once 
$m_+ \gtrsim m_T$ decays to the $W_B$ and $Z_B$ bosons 
become kinematically accessible, and begin to play an important role. 
For heavy Higgs masses we see that visible decays into $WW$, $ZZ$ and 
di-Higgs dominate, together with invisible decays. Their respective 
contributions to the branching ratio are roughly $2/7,\,1/7,\,1/7,\,3/7$, 
exactly as expected from symmetry arguments. For small $m_{+}$ we 
approach the edge of potential stability. Near the edge 
$\vartheta\sim\theta$, leading to sequestering of the two sectors. 
However, this tuned region does not correspond to a Twin Higgs-like 
potential since it is not approximately SU(4) symmetric, as we show in 
Sec. \ref{ss.recognize}.

Clearly, the potential in Eq.~\eqref{e.higgspot} is defined by four 
parameters. The measured values of $v_\text{EW}$ and $m_h=m_{-}$ already 
constrain this system. Measuring deviations in the Higgs couplings to SM 
fields determines $\cos(\vartheta-\theta)$. Currently, the LHC has 
measured some Higgs couplings to $\sim10\%$ accuracy~\cite{Khachatryan:2016vau}, and is expected to 
reach $\sim5\%$ precision by the end of the high luminosity 
run~\cite{Dawson:2013bba}. Linear electron positron colliders can reduce the 
uncertainty to better than $1\%$. Measuring the mass of the twin sector 
Higgs $m_{+}$ would then completely determine the potential. The width 
and branching ratios of the heavy Higgs are then completely 
specified. Therefore, with the mass in hand, measuring the twin sector 
Higgs rate into one or more visible states constitutes a powerful test of 
the twin Higgs framework.

Our discussion till now has focused on the case in which the discrete 
$Z_2$ symmetry is only softly broken. However, a small hard breaking of 
the discrete symmetry by the twin sector Yukawa couplings allows a 
simple resolution of the cosmological problems associated with the MTH 
framework. We therefore briefly consider the implications of a hard 
breaking of the discrete symmetry in the Yukawa sector. 

\begin{figure}[th]
\centering
\includegraphics[width=0.9\textwidth]{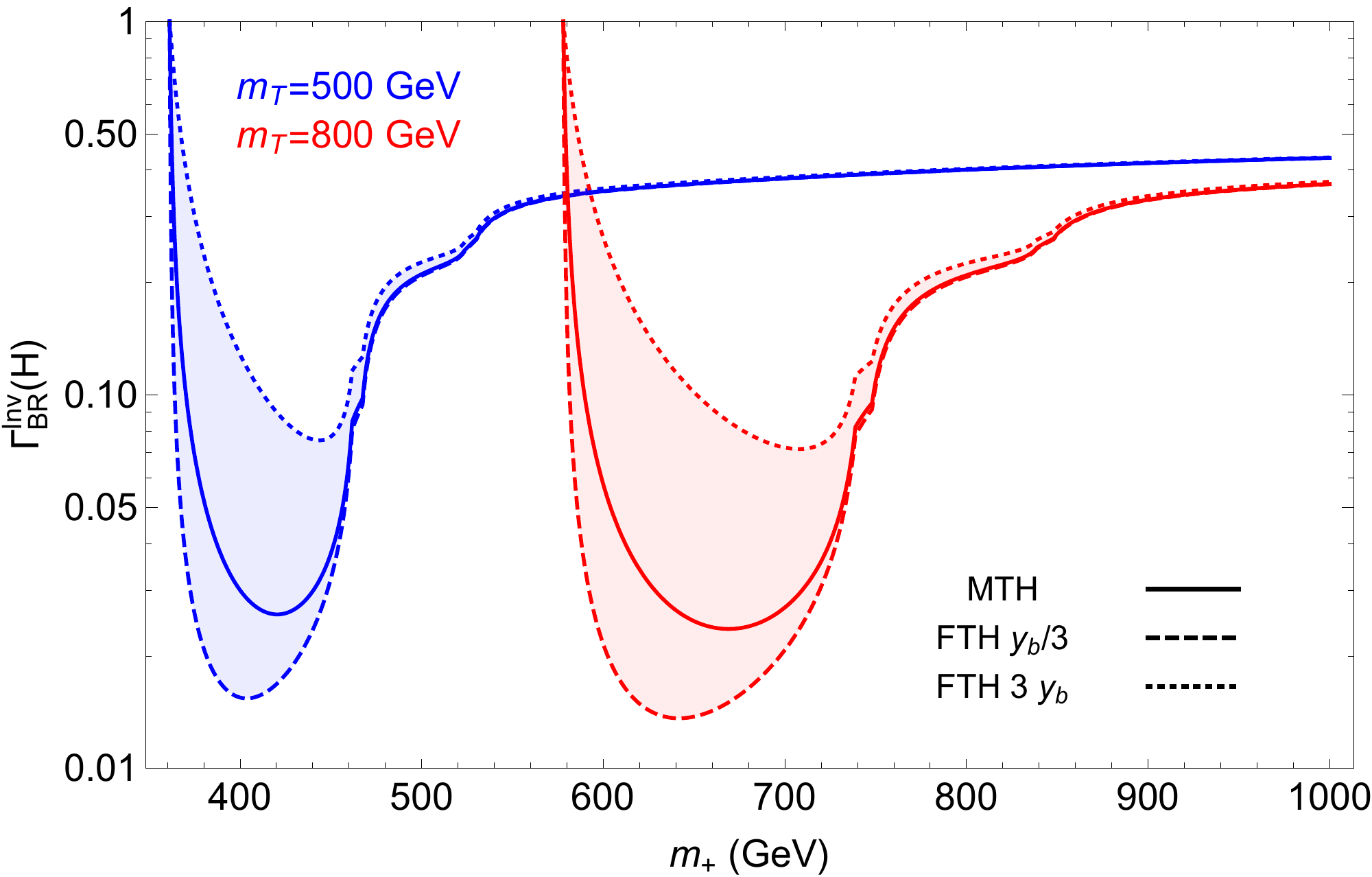}
\caption{\label{f.HiggInvisWidth} Branching fraction of the twin sector
Higgs to twin states as a function of its mass
for the MTH (solid) and FTH (dashed and dotted) models. The blue (red) curves
correspond to a mirror top mass of 500 (1000) GeV. In the FTH model the
twin bottom Yukawa varies from one third the SM value (dashed line) to
three times the SM value (dotted line). }
\end{figure}

Once the twin sector Yukawa couplings are allowed to 
vary, the invisible decay widths of both the light Higgs and the heavy Higgs are 
affected. In the case of the light Higgs, by measuring the total rate 
into both visible and invisible final states, it is still possible to 
extract $\cos(\vartheta-\theta)$. In the case of the twin sector Higgs, 
however,  
without any knowledge of the branching ratio into the 
twin sector, it is no longer possible to predict the rate into visible 
states. However, for large twin sector Higgs masses, unless the hard 
breaking of the discrete symmetry by the Yukawas is very large, the 
primary decay modes are expected to be the same as in the soft breaking 
case. This can be seen in Fig.~\ref{f.HiggInvisWidth}, where the 
invisible branching ratio of the twin sector Higgs in the FTH model has 
been plotted against $m_+$ for two different values of the twin bottom 
Yukawa coupling and compared against the invisible branching ratio in 
the MTH model. We see that once $m_+ > m_T$, the different curves 
quickly converge towards $3/7$, the theoretical prediction. We conclude 
that for heavy twin sector Higgs bosons, if the hard breaking of the 
discrete symmetry is small, the predictions of the MTH continue to hold to
a good approximation.
 
\subsection{Fine Tuning in the Model \label{ss.recognize}}

In the Twin Higgs framework, the global symmetry of the Higgs sector 
is ultimately what protects the mass of the light Higgs from large 
radiative corrections. Therefore, the SU(4) violating parameters $m^2$ 
and $\delta$ in the Higgs potential, Eq. \eqref{e.higgspot}, must be 
small compared to their SU(4) invariant counterparts, $\mu^2$ and 
$\lambda$, in order 
for the visible sector Higgs to be naturally light. 
Dividing Eq. \eqref{e.mCondition} by Eq. \eqref{e.muCondition} we obtain
 \begin{equation}
\frac{m^2}{\mu^2}=\frac{\frac{\delta}{\lambda}\cos2\vartheta}{2+\frac{\delta}{\lambda}},
\label{e.mOverMu}
 \end{equation}
 which shows that as long as $\delta\ll \lambda$ the potential is 
approximately SU(4) symmetric, and corresponds to a Twin Higgs potential.

\begin{figure}[th]
\centering\includegraphics[width=0.6\textwidth]{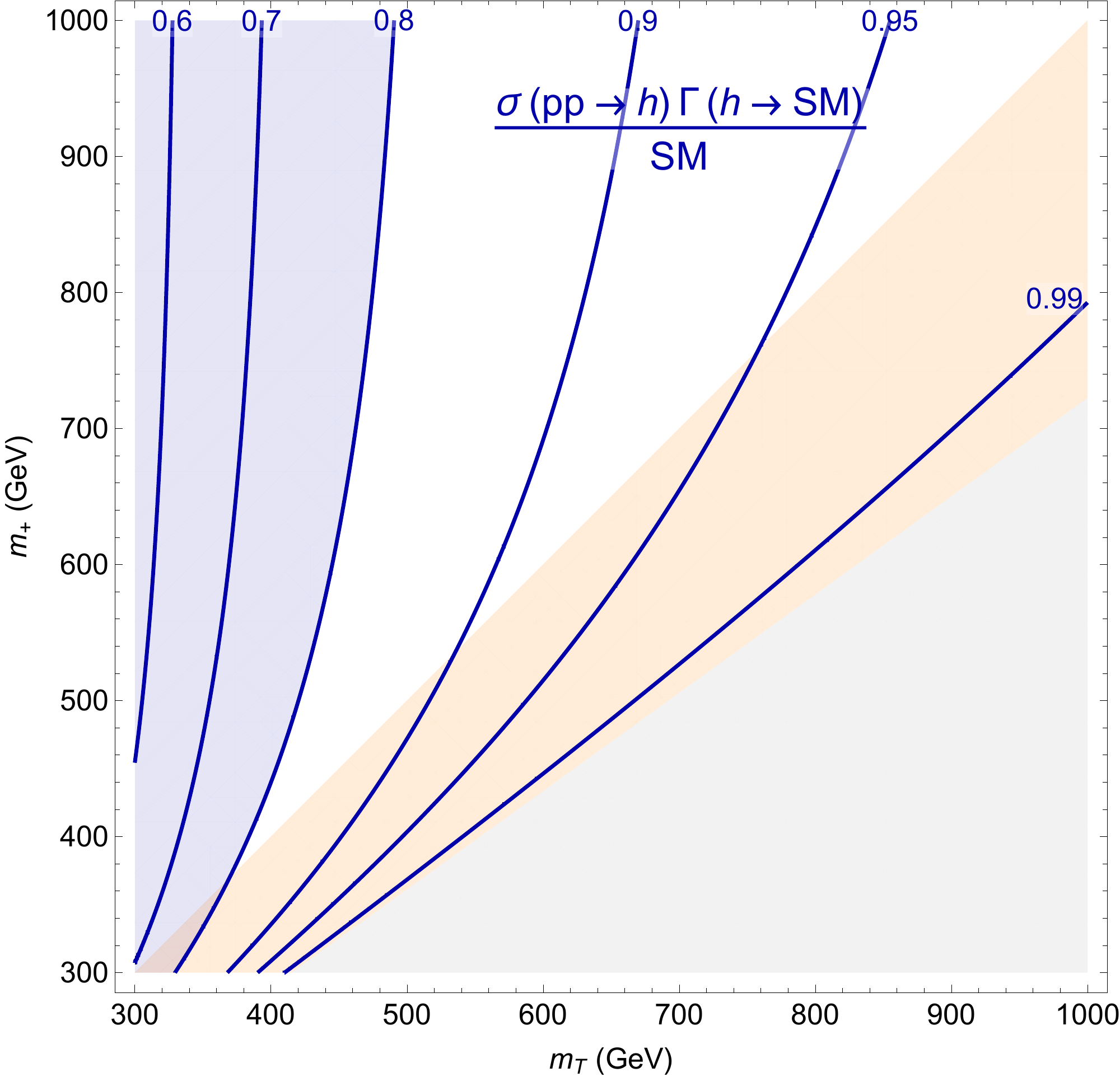}
\caption{\label{f.HiggsCouplings} The region with stable vacuum but with $m_+<m_T$, and hence the tuning is \emph{not} significantly improved over the SM, is shaded orange. The grey region does not allow a stable vacuum. The blue shaded region is disfavored by LHC Higgs coupling measurements. }
\end{figure}

To determine the fine-tuning in this model, note that the parameter $\mu^2$ 
receives radiative corrections from the top loop, just as the Higgs mass 
parameter in the SM does. Assuming a cutoff $\Lambda$, the fine-tuning 
associated with the sensitivity of $\mu^2$ to the cutoff is given by
 \begin{equation}
\frac{3 Y_t^2}{8 \pi^2} \frac{\Lambda^2}{\mu^2} \; .
 \end{equation}
 If the potential is approximately SU(4) invariant, so that $\lambda \gg 
\delta$, we can combine Eqs.~\eqref{e.muCondition} and 
\eqref{masseigenvalues} to obtain $2 \mu^2 \simeq m_+^2$. Then the 
expression for this fine-tuning reduces to
 \begin{equation}
\frac{3 Y_t^2}{4 \pi^2} \frac{\Lambda^2}{m_+^2} \; .
\label{FT1}
 \end{equation}
 Now, it follows from Eqs.~\eqref{e.mOverMu} and \eqref{e.muCondition} 
that the electroweak scale $v_{\rm EW}$ is related to $\mu^2$ as
 \begin{equation}
 v_{\rm EW}^2 = \frac{\mu^2}{2 \lambda + \delta} - \frac{m^2}{\delta} \; .
 \end{equation}
 Then the sensitivity of the electroweak scale to the parameter $\mu^2$ is
given by
 \begin{equation}
\frac{\mu^2}{v_{\rm EW}^2}
\frac{\partial \; v_{\rm EW}^2}{\partial \; \mu^2}    
 = \frac{\mu^2}{v_\text{EW}^2\left(2\lambda + \delta\right)} = \frac{f^2}{v_{\rm EW}^2} 
 \simeq \frac{m_T^2}{2 m_t^2} \;.
 \label{FT2}
 \end{equation}
 The sensitivity of the electroweak scale to the cutoff in this model is 
obtained from Eqs.~\eqref{FT1} and \eqref{FT2} as
 \begin{equation}
\frac{3 Y_t^2}{4 \pi^2} \frac{\Lambda^2}{m_+^2} \frac{m_T^2}{2m_t^2} \; .
\label{FTMTH}
 \end{equation}
 In comparison, the sensitivity of the electroweak scale to the cutoff 
in the SM is given by
 \begin{equation}
\frac{3 Y_t^2}{4 \pi^2} \frac{\Lambda^2}{m_-^2} \; ,
\label{FTSM}
 \end{equation}
 where $m_-$ is 125 GeV, the mass of the light Higgs. Then the improvement 
in fine-tuning with respect to the SM for the same cutoff is obtained by 
taking the ratio of Eq.~\eqref{FTSM} to Eq.~\eqref{FTMTH},
 \begin{equation}
2 \frac{m_+^2}{m_-^2} \frac{m_t^2}{m_T^2} \; .
 \end{equation}
 We see from this that, for a given top partner mass, a heavier twin 
sector Higgs is preferred. We require $m_+ > m_T$ to obtain any 
significant improvement in fine-tuning with respect to the SM.

\section{Current and Future LHC Reach\label{s.LHC}}

As in all pNGB Higgs models, the Twin Higgs framework predicts reduced 
couplings of the light Higgs to SM states, resulting in a suppression of 
the Higgs production cross section. This, together with Higgs decays into invisible twin sector final states, results in fewer Higgs 
events. The resulting LHC constraints on the MTH model were calculated 
in \cite{Burdman:2014zta}, but this analysis assumed the radial 
mode was heavy, with mass near the cutoff of the model. Here we 
determine the suppression of Higgs rates relative to the SM, taking into 
account the finite mass of the radial mode. In 
Fig.~\ref{f.HiggsCouplings} we display contours of the ratio of Higgs 
event rates to visible sector states in the MTH model relative to Higgs 
event rates in the SM. We see that for lighter radial modes the 
deviation from the SM value decreases.

The LHC has already measured the one sigma Higgs couplings to EW gauge bosons to 
$\sim 10\%$~\cite{Khachatryan:2016vau}. This implies that the region to the left of the 0.8 contour 
is already ruled out. The HL-LHC is expected to probe up to the $0.9$ 
contour. As the figure shows, the LHC Higgs coupling measurements cannot 
fully probe the parameter space of the MTH model.

\begin{figure}[th]
\centering
\includegraphics[width=0.60\textwidth]{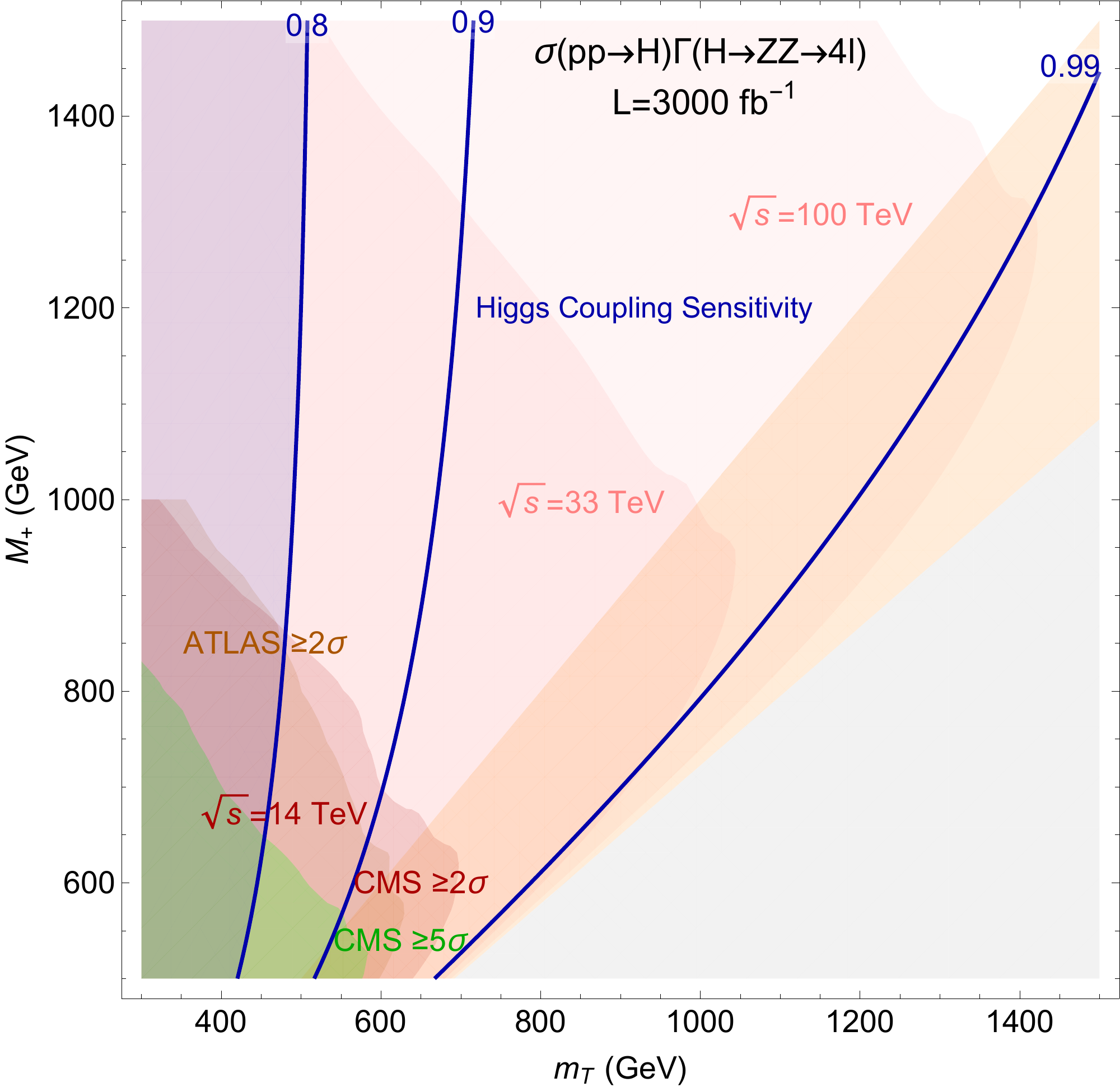}
\caption{\label{f.HLLHC} Exclusion bound on the Twin Higgs  model in the mass plane of twin top and twin sector Higgs for the High Luminosity LHC run, as well as for future hadron colliders with center of mass energies of 33 and 100~TeV. The extrapolation for the HL-LHC is made by ATLAS and CMS for the $H\to ZZ\to 4\ell$ process. Blue contours denote variation in Higgs couplings, with the region to the left of 0.8 already excluded by LHC measurements. The orange region does not significantly improve tuning compared to the SM. The extrapolation to future hadron colliders is estimated using Fig. 3 of Ref.~\cite{Buttazzo:2015bka}. See text for further details.}
\end{figure}

The LHC has a strong experimental program directly searching for heavy 
Higgs-like scalars $H$, but present search limits are weaker than Higgs coupling constraints.
However, both ATLAS~\cite{ATL-PHYS-PUB-2013-016} and CMS~\cite{CMS:2013dga} have 
estimated the high luminosity reach for a heavy scalar in the $H\to 
ZZ\to 4\ell$ channel. In Fig. \ref{f.HLLHC} we see the exclusion curves 
for both detectors as well as the discovery region for CMS. The reach 
for ATLAS is expected to be comparable. We use the ATLAS exclusion 
numbers as well as the CMS exclusion and 5$\sigma$ numbers from Fig. 3 
of \cite{Holzner:2014qqs}.  We translate the bounds into our framework for any chosen masses of the twin sector Higgs and twin sector top by first rescaling the SM Higgs cross section provided by the LHC Higgs Cross section Working Group~\cite{deFlorian:2016spz,HWG} by the factor of $\sin^{2}(\vartheta-\theta)$. Then using the results collected in the Appendix we include the branching fraction of $h_+$ into $ZZ$. This leads to the 
red and green shaded contours in Fig.~\ref{f.HLLHC}. These searches 
can probe twin sector Higgs masses as heavy as a TeV, but only if the 
twin top is light. Unfortunately, the greater part of this region is 
already excluded by Higgs coupling measurements. It is only in a limited 
region of allowed parameter space that the LHC will be able to discover 
the twin sector Higgs.

If Higgs coupling deviations are observed at the HL-LHC and a heavy 
scalar is found in the four lepton search, the entire Higgs potential 
will be specified. This leads to a specific prediction for the $H\to ZZ$ 
rate, which can be compared with the experimental results, explicitly 
testing the mechanism. Unfortunately, the region of parameter space in 
which such a scenario can play out is rather limited.

For completeness we also include in Fig.~\ref{f.HLLHC} an estimate for the sensitivity at future hadron colliders, with center of mass energies of 33 and 100~TeV. In order 
to estimate the signal cross section in our model at future colliders, we scale up the cross section for the twin sector Higgs at 14~TeV (obtained as described above) by 
the appropriate ratio of PDF luminosities, using Fig. 3 of Ref.~\cite{PDFratio} for 33~TeV, and Table 2 of Ref.~\cite{Mangano:2016jyj} for 100~TeV. We then compare the 
cross section (times branching ratio) thus obtained to Fig. 3 of Ref.~\cite{Buttazzo:2015bka}, where the sensitivity at these colliders to heavy scalars decaying to $ZZ$ 
was estimated, under the assumption that the background is primarily $q$-$\bar{q}$ initiated. As can be seen from Fig.~\ref{f.HLLHC}, going to higher energy significantly 
extends the region of parameter space where hadron colliders have sensitivity to the twin sector Higgs.


\section{Linear Collider Reach\label{s.LinCol}}

In this section, we discuss the potential for the next generation of 
linear colliders to discover the twin sector Higgs, and determine the 
parameters in the scalar sector of the MTH model. For concreteness, we 
focus on the ILC and CLIC proposals. These colliders possess two 
advantages with respect to the LHC.
 \begin{itemize} 
 \item{Both of these machines will be able to measure the couplings of 
the light Higgs to better than $1\%$, which, as can be seen from Fig.~\ref{f.HiggsCouplings}, covers most of the preferred parameter space.}
 \item{Because of their much lower backgrounds, these colliders 
potentially have much greater reach for the twin sector Higgs.}
 \end{itemize}
 As explained earlier, the mass of the twin sector Higgs, together with 
measurements of the deviations in couplings of the light Higgs, 
completely specifies the scalar potential of the MTH model. Then, a 
measurement of the rate of twin sector Higgs events into any SM final 
state overdetermines the scalar potential, and constitutes a powerful 
test of this framework.

For our analysis, we focus on benchmark scenarios motivated by the ILC 
and CLIC proposals. For the high energy ILC, which is a 1 TeV machine, 
we consider two benchmark scenarios corresponding to 1 $\text{ab}^{-1}$ 
and 3 $\text{ab}^{-1}$ of integrated luminosity. The CLIC benchmark 
corresponds to a 1.5 TeV machine with an integrated luminosity of 1.5 
$\text{ab}^{-1}$. Signals are generated in {\sc MadGraph5} 
\cite{Alwall:2014hca} and showered with {\sc Pythia8} 
\cite{Sjostrand:2014zea}. We use the {\sc Delphes3} 
\cite{deFavereau:2013fsa} detector simulator with the anti-$k_\text{T}$ 
clustering algorithm \cite{Cacciari:2008gp} and {\sc FastJet} 
\cite{Cacciari:2011ma} library to simulate the detector. We use the {\sc 
Delphes} card based on the ILC construction outlined in 
\cite{Behnke:2013lya}. A simulation of the CLIC detector is not yet 
available, but is expected to be qualitatively similar.

The branching ratios shown in Fig. \ref{f.TwinHiggsBR} show that of the 
twin sector Higgs' visible decays products, $WW$ is the largest. 
However, the $WW$ background is prohibitively large. Instead we focus on 
decays to di-Higgs, for which the background is orders of magnitude 
smaller, making the process $h_+\to h_- h_-\to 4b$ very attractive. 
While the 4$b$ final state is difficult to extract from background at a 
hadron collider, the comparatively clean environment of a lepton machine 
is admirably suited to such a search.

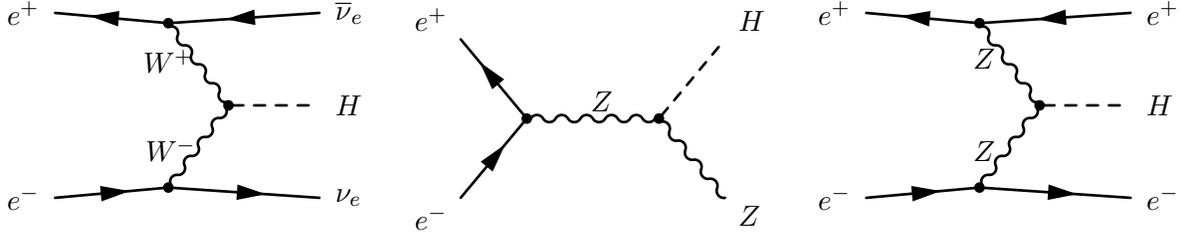
\begin{figure}
\begin{subfigure}[b]{0.32\textwidth}
\begin{fmffile}{Wfusion}
\begin{fmfgraph*}(100,70)
\fmfpen{1.0}
\fmfstraight
\fmfleft{i1,p2,i3}\fmfright{o1,o2,o3}
\fmfv{l= $e^-$,l.a=180}{i1}\fmfv{l=$e^+$,l.a=180}{i3}\fmfv{l=$\nu_e$,l.a=0}{o1}\fmfv{l=$\overline{\nu}_e$,l.a=0}{o3}\fmfv{l=$H$}{o2}
\fmf{fermion,tension=1.2}{i1,v1}\fmf{fermion,tension=0.8}{v1,o1}
\fmf{fermion,tension=1.2}{v3,i3}\fmf{fermion,tension=0.8}{o3,v3}
\fmf{boson,tension=0.25}{v1,v2}
\fmf{boson,tension=0.25}{v3,v2}
\fmf{dashes, tension=0.33}{v2,o2}
\fmfv{decor.shape=circle,decor.filled=full,decor.size=1.5thick,l=$W^-$,l.a=90,l.d=10}{v1} 
\fmfv{decor.shape=circle,decor.filled=full,decor.size=1.5thick}{v2} 
\fmfv{decor.shape=circle,decor.filled=full,decor.size=1.5thick,l=$W^+$,l.a=-90,l.d=10}{v3} 
\end{fmfgraph*}
\end{fmffile}
\end{subfigure}
\begin{subfigure}[b]{0.32\textwidth}
\begin{fmffile}{AProd}
\begin{fmfgraph*}(100,60)
\fmfpen{1.0}
\fmfstraight
\fmfleft{i1,p1,i2}\fmfright{o1,p2,o2}
\fmfv{l= $e^-$}{i1}\fmfv{l=$e^+$}{i2}\fmfv{l=$Z$}{o1}\fmfv{l=$H$}{o2}
\fmf{fermion,tension=1}{i1,v1,i2}
\fmf{boson,tension=1}{v1,v2,o1}
\fmf{dashes, tension=1}{v2,o2}
\fmfv{decor.shape=circle,decor.filled=full,decor.size=1.5thick,l=$Z$,l.a=15,l.d=25}{v1} 
\fmfv{decor.shape=circle,decor.filled=full,decor.size=1.5thick}{v2} 
\end{fmfgraph*}
\end{fmffile}
\end{subfigure}
\begin{subfigure}[b]{0.32\textwidth}
\begin{fmffile}{Zfusion}
\begin{fmfgraph*}(100,70)
\fmfpen{1.0}
\fmfstraight
\fmfleft{i1,p2,i3}\fmfright{o1,o2,o3}
\fmfv{l= $e^-$,l.a=180}{i1}\fmfv{l=$e^+$,l.a=180}{i3}\fmfv{l=$e^-$,l.a=0}{o1}\fmfv{l=$e^+$,l.a=0}{o3}\fmfv{l=$H$}{o2}
\fmf{fermion,tension=1.2}{i1,v1}\fmf{fermion,tension=0.8}{v1,o1}
\fmf{fermion,tension=1.2}{v3,i3}\fmf{fermion,tension=0.8}{o3,v3}
\fmf{boson,tension=0.25}{v1,v2}
\fmf{boson,tension=0.25}{v3,v2}
\fmf{dashes, tension=0.33}{v2,o2}
\fmfv{decor.shape=circle,decor.filled=full,decor.size=1.5thick,l=$Z$,l.a=80,l.d=10}{v1} 
\fmfv{decor.shape=circle,decor.filled=full,decor.size=1.5thick}{v2} 
\fmfv{decor.shape=circle,decor.filled=full,decor.size=1.5thick,l=$Z$,l.a=-80,l.d=10}{v3} 
\end{fmfgraph*}
\end{fmffile}
\end{subfigure}
\caption{\label{f.production}Dominant Higgs production mechanisms at lepton colliders.}
\end{figure}

\begin{figure}[th]
\centering
\begin{subfigure}[b]{0.49\textwidth}
\includegraphics[width=\textwidth]{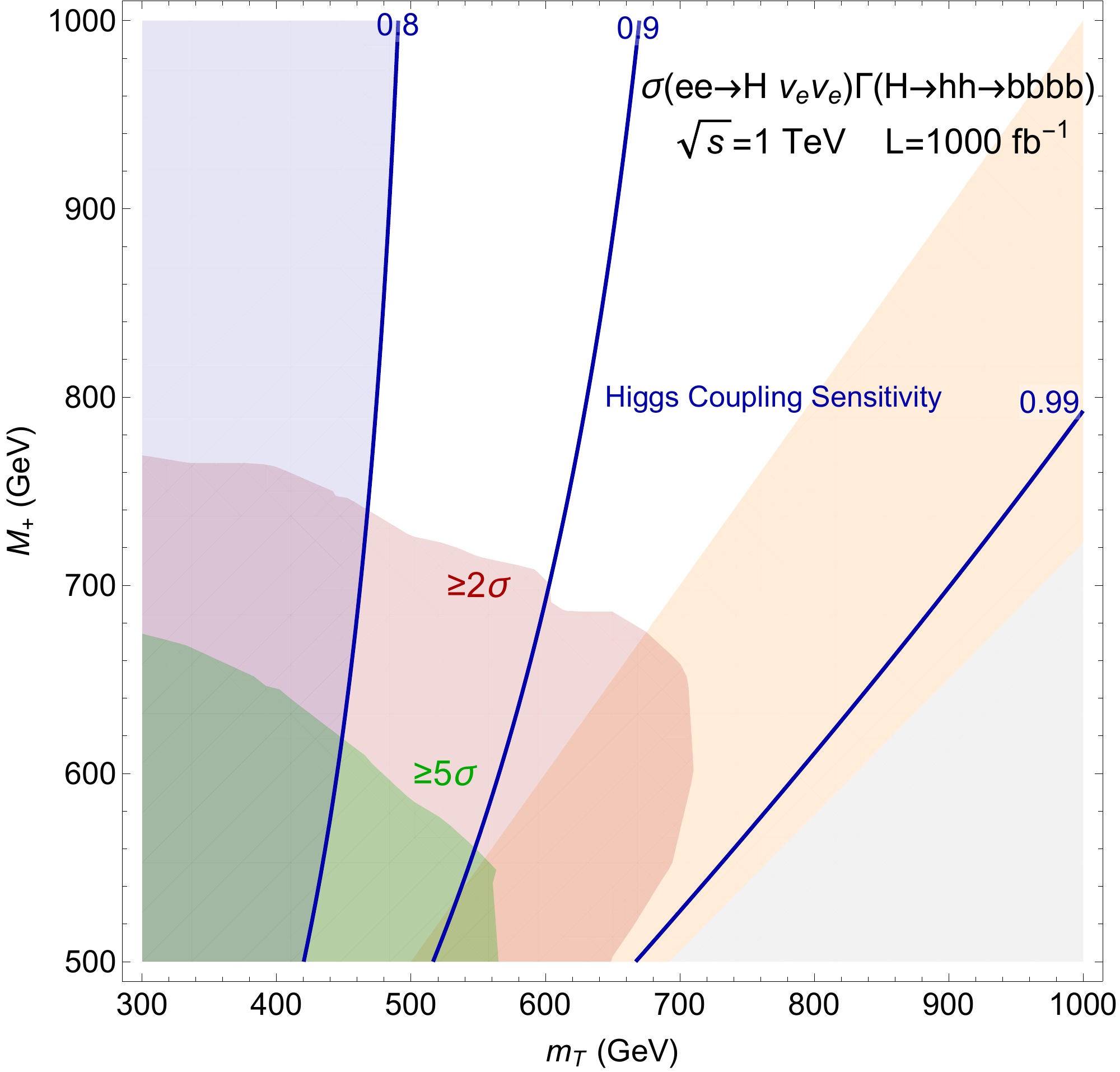}
\end{subfigure}
\begin{subfigure}[b]{0.49\textwidth}
\includegraphics[width=\textwidth]{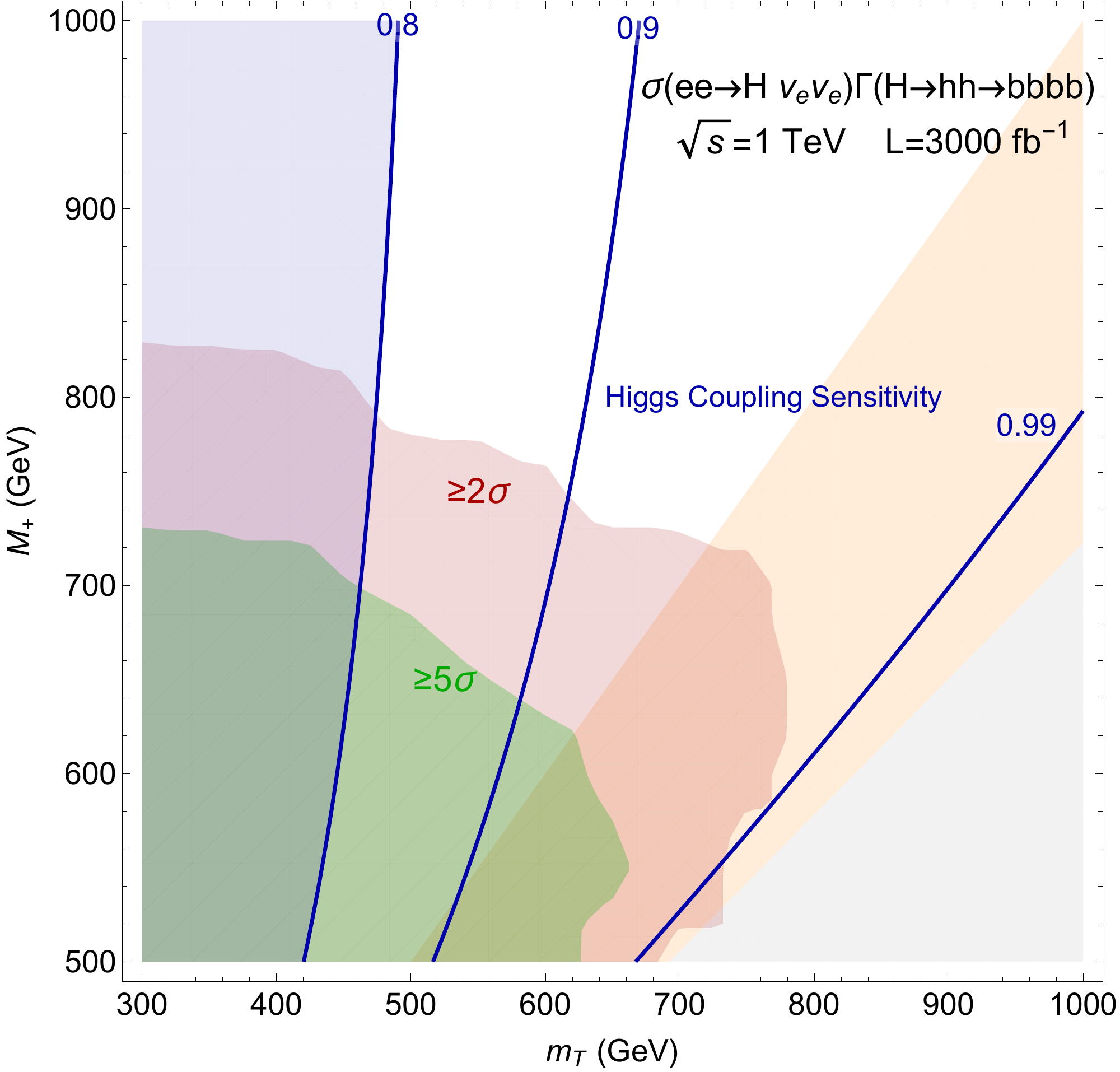}
\end{subfigure}
\caption{\label{f.WWfusionResults} Results for both the ILC 1 $\text{ab}^{-1}$ (left) and 3 $\text{ab}^{-1}$ (right) benchmark linear collider scenarios for $W$ fusion to heavy twin sector Higgs decaying to di-Higgs to 4 $b$s. As in Fig. \ref{f.HLLHC}, the blue contours indicate deviation in Higgs couplings, the region to the left of 0.8 excluded by current measurements. The gray region does not provide a stable vacuum. }
\end{figure}

\begin{figure}[th]
\centering
\includegraphics[width=0.65\textwidth]{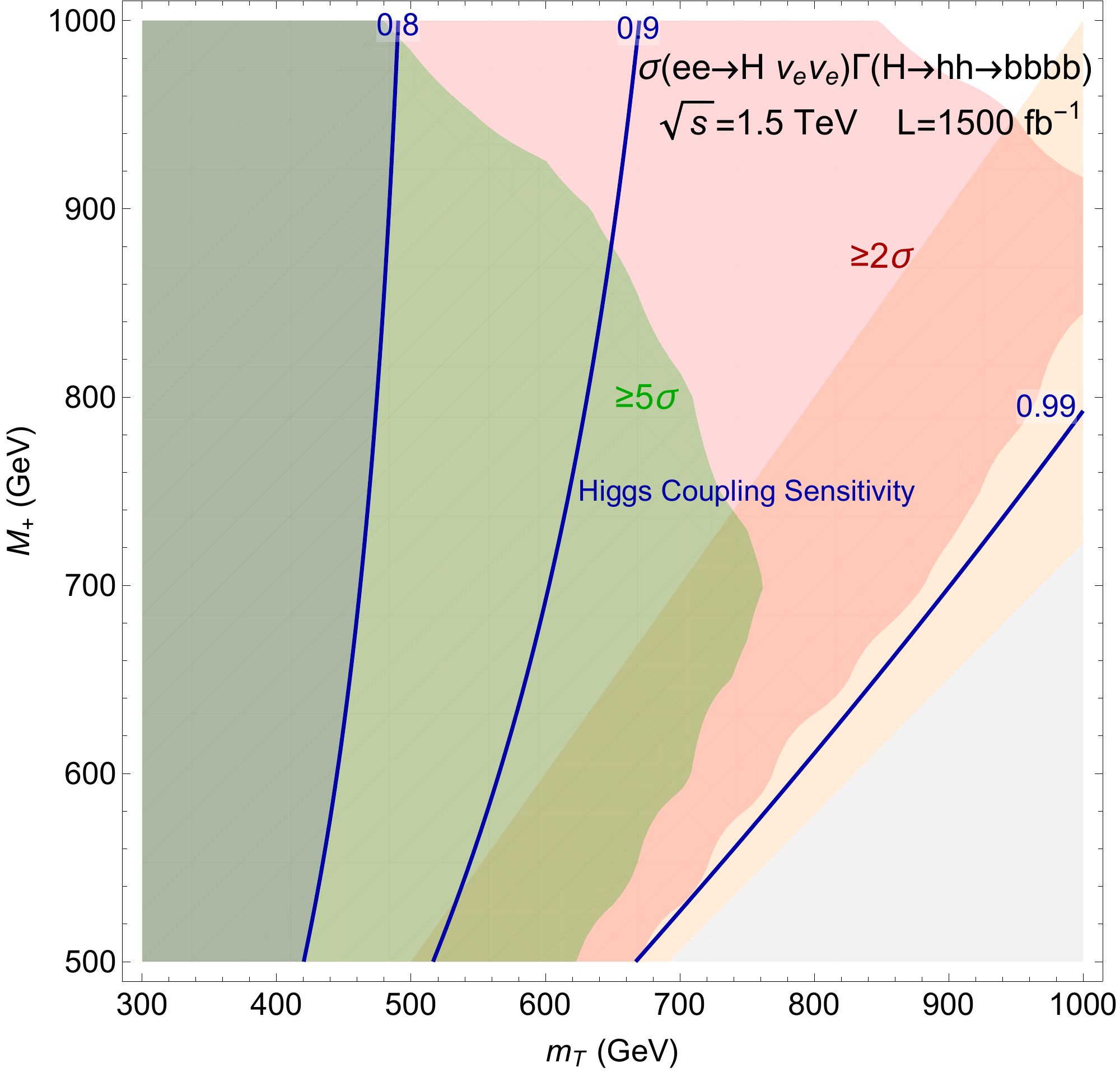}
\caption{\label{f.WWfusionCLIC} Results for both the CLIC benchmark linear collider scenario for $W$ fusion to heavy twin sector Higgs decaying to di-Higgs to 4 $b$s. As in Fig. \ref{f.HLLHC}, the blue contours indicate deviation in Higgs couplings, the region to the left of 0.8 excluded by current measurements. The gray region does not provide a stable vacuum. }
\end{figure}

Figure~\ref{f.production} displays the dominant Higgs production processes 
at lepton colliders. Our analysis employs the dominant Higgs production 
process at high energies, which is $WW$ fusion. Our study required at 
least three jets, each required to have $p_T> $20 GeV and $|\eta|<2.5$. 
In addition, we demand three $b$-tags and that the jets reconstruct two 
on-shell Higgses, with on-shell window (75 GeV, 135 GeV). We considered 
invariant mass bins that contained at least 85\% of the signal events in 
the four bins surrounding the peak that passed the previous cuts. These 
bins were taken to be several tens of GeV wide to accommodate the 
expected jet energy resolution. In Fig.~\ref{f.WWfusionResults} we see 
the results for both the ILC benchmarks. 
Comparing this figure to the LHC results in Fig. \ref{f.HLLHC}, we find 
that the reach of the ILC with 1 $\text{ab}^{-1}$ is comparable to 
that of the HL-LHC. With 3 $\text{ab}^{-1}$ the ILC will be able to 
discover the twin sector Higgs for a greater range of twin top and twin 
sector Higgs masses than the LHC. The higher energy and increased 
luminosity of the CLIC benchmark allow even greater opportunity to 
discover the twin sector Higgs, as can be seen from Fig.~\ref{f.WWfusionCLIC}.

\begin{figure}
\centering
\begin{subfigure}[b]{0.49\textwidth}
\includegraphics[width=\textwidth]{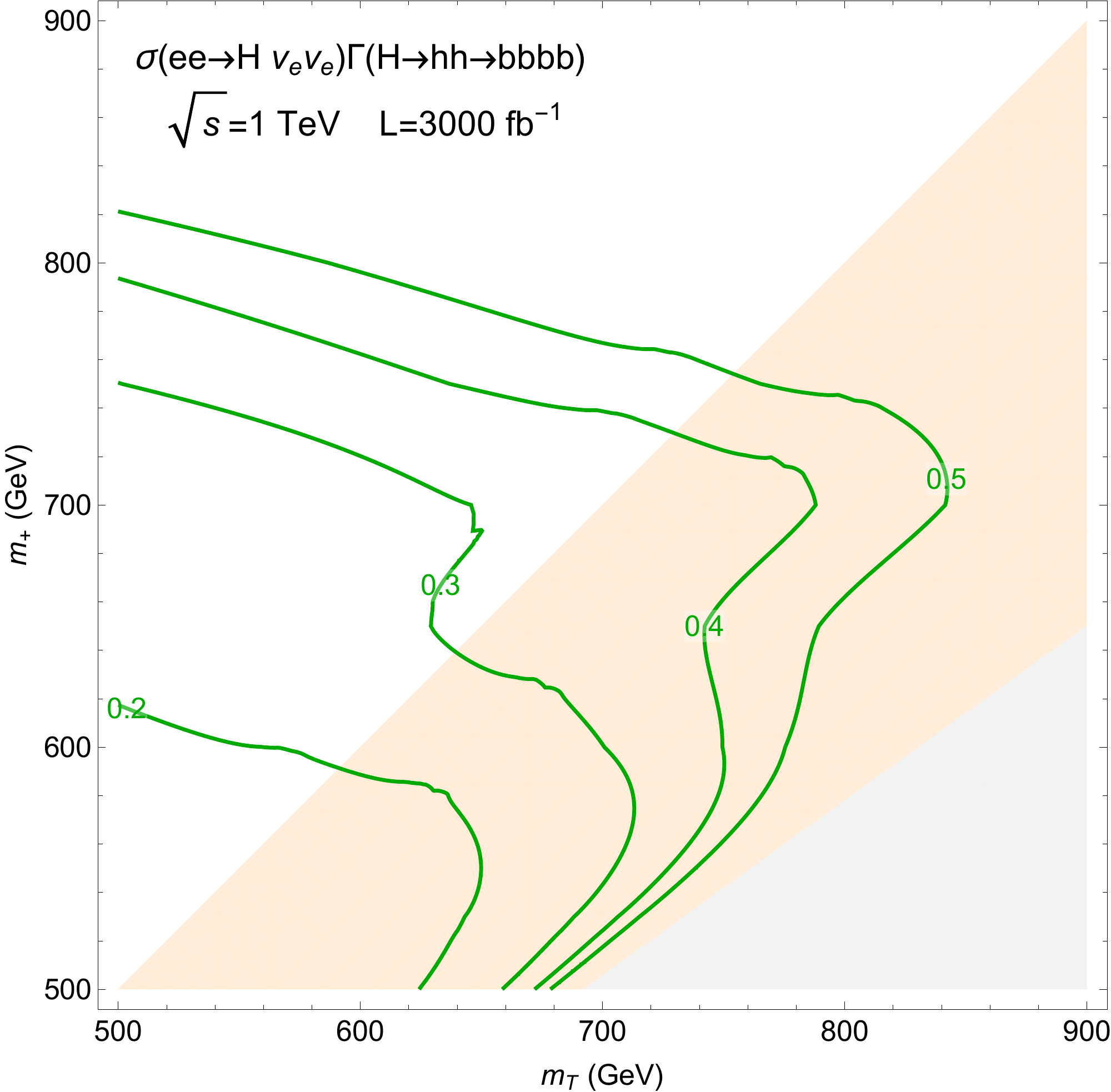}
\end{subfigure}
\begin{subfigure}[b]{0.49\textwidth}
\includegraphics[width=\textwidth]{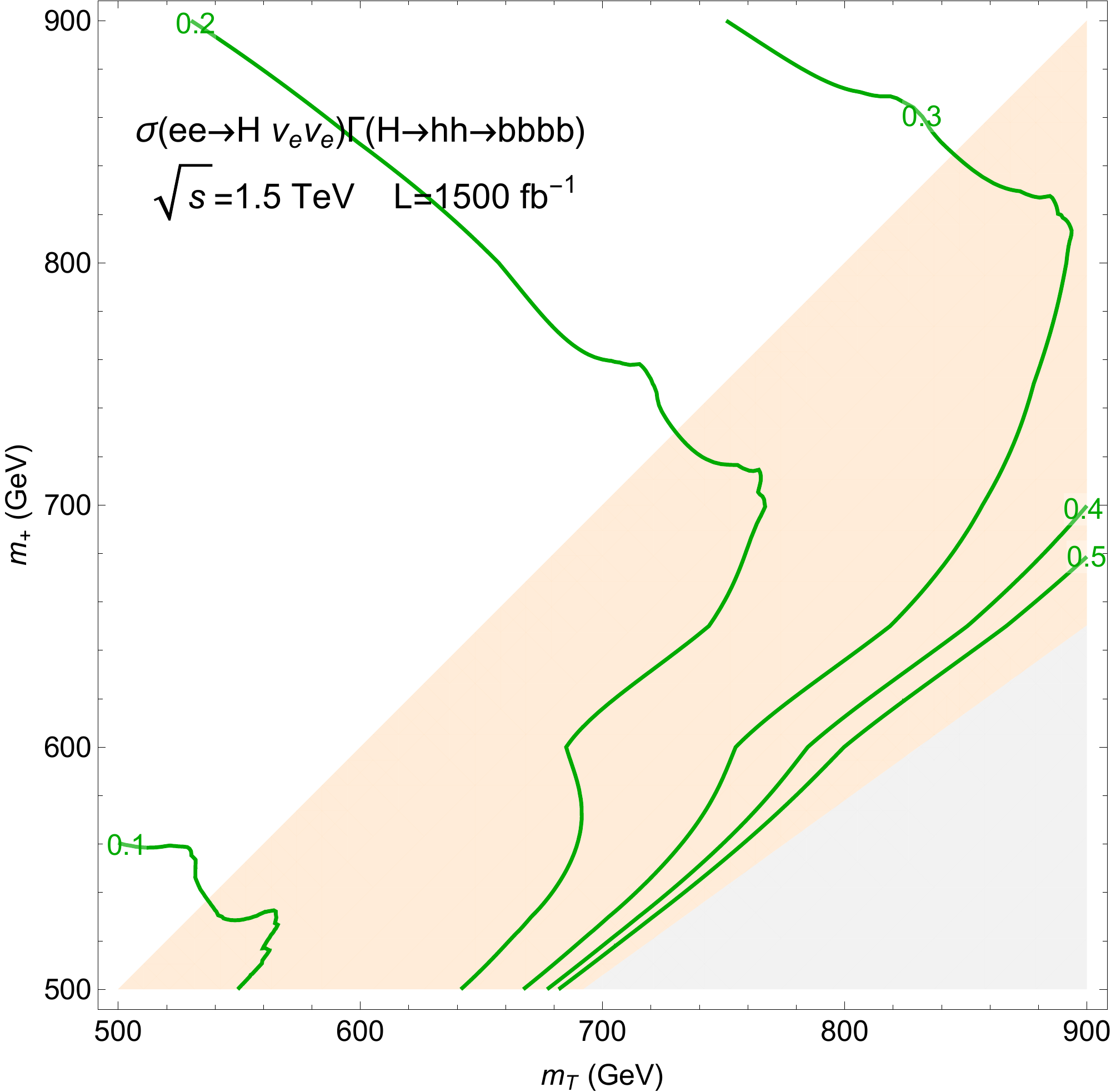}
\end{subfigure}
\caption{\label{f.Precision} The uncertainty in the ratio of observed and expected events, centered around the value 1, for the ILC ($\int\! dt\mathcal{L}=3~\text{ab}^{-1}$) (left) and CLIC ($\int\! dt\mathcal{L}=1.5~\text{ab}^{-1}$) (right) benchmarks as a function of $m_{+}$ and $m_{T}$. See main text for additional details.}
\end{figure}

Finally, we quantify the confidence with which the Twin Higgs mechanism can be confirmed as follows. For a given parameter point, we calculate the uncertainty in the number of observed events after the cuts described above (due to Poisson statistics), and we also estimate the uncertainty in the expected number of events at that parameter point, where the leading contribution is the uncertainty in the value of $\sin^{2}(\vartheta-\theta)$ arising from Higgs coupling measurements. In particular, we assume that $\kappa_{Z}$, the multiplicative factor that measures the deviation of the Higgs coupling to the $Z$-boson, can be measured with a precision of 0.5\%~\cite{Dawson:2013bba}. Combining the uncertainties in the number of expected and observed events, we arrive at the fractional uncertainty in the ratio of observed to expected events, centered around the value 1. The fractional uncertainty is plotted for the ILC ($\int\! dt\mathcal{L}=3~{\rm ab}^{-1}$) and CLIC (${\int\! dt\mathcal{L}}=1.5~{\rm ab}^{-1}$) benchmarks in Fig.~\ref{f.Precision}.

\section{Conclusion\label{s.conclusion}}

The Twin Higgs mechanism protects the mass of the Higgs against 
radiative corrections without requiring new particles charged under the 
SM gauge groups. In this framework, the light Higgs emerges as the pNGB 
associated with the breaking of a global symmetry, and its mass is 
protected against quantum effects by a combination of the global 
symmetry and a discrete $Z_2$ symmetry. If the breaking of the global 
symmetry is realized linearly, the radial mode of the Higgs potential, 
the twin sector Higgs, is present in the spectrum. This particle 
provides a new portal between the visible and twin sectors. We have 
shown that, if the discrete $Z_2$ symmetry is only softly broken, a 
measurement of the mass of the twin sector Higgs, when combined with 
precision measurements of the light Higgs, completely 
specifies the Higgs potential. The rates for twin sector Higgs events 
are then testable predictions of the Twin Higgs framework. This 
conclusion also applies to theories that exhibit hard breaking of the 
$Z_2$ symmetry by the twin sector Yukawa couplings, provided that this 
breaking is small enough that the correction to the overall width of the 
twin sector Higgs is small. While the high luminosity LHC can 
potentially discover the twin sector Higgs, linear colliders such as the 
ILC or CLIC have much better precision and greater reach, allowing them 
to test the Twin Higgs framework.

\acknowledgments
 We thank Jessie Shelton for helpful communication. SN would like to 
thank Aqeel Ahmed for useful discussions. ZC is supported in part by the 
National Science Foundation under Grant Number PHY-1620074, the Fermilab 
Intensity Frontier Fellowship and the Visiting Scholars Award \#17-S-02 
from the Universities Research Association. CK is supported by the 
National Science Foundation under Grant Number PHY-1620610. CBV is 
supported by Department of Energy Grant Number DE-SC-0009999. ZC would 
like to thank the Fermilab Theory Group for hospitality during the 
completion of this work. CK would also like to thank the Aspen Center 
for Physics, which is supported by National Science Foundation grant 
PHY-1607611, where part of this work was completed.

\appendix

\section{Higgs Sector Couplings and Decays\label{a.Decays}}

This appendix includes several formulae relevant to the production and 
decays of Higgs bosons.

\subsection{Loop Induced Couplings of the Higgs Fields}

 Decays of the Higgs bosons to photons and gluons (we neglect the 
subdominant $Z\gamma$ channel) proceed through loops of electrically 
charged or colored particles. The expressions for these decay rates 
employ the functions
 \begin{align} 
A_F(x)&=2x^2\left[\frac{1}{x}+\left(\frac{1}{x}-1 \right)f(x) \right],\\ 
A_V(x)&=-x^2\left[\frac{2}{x^2}+\frac{3}{x}+3\left(\frac{2}{x}-1\right)f(x) 
\right], 
 \end{align} 
 with 
 \begin{equation} 
f(x)=\left\{\begin{array}{lr} \arcsin^2\left(\frac{1}{\sqrt{x}}\right) & x\geq 1\\ 
-\frac14\left[\ln\left(\frac{1+\sqrt{1-x}}{1-\sqrt{1-x}} \right)-i\pi 
\right]^2 & x<1 \end{array}\right. . 
 \end{equation} 
 The loop induced coupling of $h_{-}$ to SM photons and gluons are 
reduced by a factor of $\cos(\vartheta-\theta)$, just like the tree 
level couplings. This is because each diagram involves only a single 
coupling of the Higgs to the particles running in the loop. This uniform 
reduction factors out of the overall amplitude. 

The coupling of $h_{+}$ to SM photons and gluons differs from the 
corresponding SM coupling in two ways. First, there is a uniform reduction 
of the tree level couplings by $\sin(\vartheta-\theta)$ that factors out of 
the amplitude, just as in the case of the light Higgs. Second, the loop 
functions $A_{F,V}(x)$ depend on the mass of the decaying particle 
through
 \begin{equation} 
x=4\frac{m_{F,V}^2}{m_\pm^2}. 
 \end{equation} 
 As the mass of the twin sector Higgs increases, $x$ decreases. Then the 
expressions for the decay width of the heavy Higgs to visible sector 
photons or gluons are the same as in the SM for a Higgs \emph{of mass} $m_{+}$ 
multiplied by $\sin^2(\vartheta-\theta)$.
 
 For instance, consider the decays of $h_{-}$ and $h_{+}$ to SM photons. 
To leading order, they are given by (see for instance \cite{Djouadi:2005gi}),
 \begin{align}
\Gamma(h_{-}\to \gamma_A\gamma_A)&=\frac{\alpha_A^2 m_{-}^3}{256 \pi^3 v_\text{EW}^2}\left| A_V\left( \frac{4m_{W,A}^2}{m_{-}^2}\right)+\sum_{f_A}N_cQ_f^2A_F\left( \frac{4m_{f,A}^2}{m_{-}^2}\right)\right|^2\cos^2(\vartheta-\theta),\\
\Gamma(h_{+}\to \gamma_A\gamma_A)&=\frac{\alpha_A^2 m_{+}^3}{256 \pi^3 v_\text{EW}^2}\left| A_V\left( \frac{4m_{W,A}^2}{m_{+}^2}\right)+\sum_{f_A}N_cQ_f^2A_F\left( \frac{4m_{f,A}^2}{m_{+}^2}\right)\right|^2\sin^2(\vartheta-\theta),
 \end{align}
 where $Q_f$ are the electric charges of the fermions and $N_c$ the 
number of colors of the various fermions. The corresponding expressions 
for decays to $B$-sector photons have exactly the same form, but now depend on 
the masses of the $B$ gauge bosons and fermions, while the 
$\cos(\vartheta-\theta)$ and $\sin(\vartheta-\theta)$ factors are 
exchanged. The form is illustrated by the leading order Higgs decays 
into $B$-sector gluons,
 \begin{align}
\Gamma(h_{-}\to g_Bg_B)&=\frac{\alpha_{s,B}^2m_{-}^3}{72\pi^3v_B^2}\left| \frac34 \sum_{q_B} A_F\left( \frac{4m_{q,B}^2}{m_{-}^2}\right) \right|^2\sin^2(\vartheta-\theta),\\
\Gamma(h_{+}\to g_Bg_B)&=\frac{\alpha_{s,B}^2m_{+}^3}{72\pi^3v_B^2}\left| \frac34 \sum_{q_B} A_F\left( \frac{4m_{q,B}^2}{m_{+}^2}\right) \right|^2\cos^2(\vartheta-\theta).
 \end{align}

 The production of Higgs bosons is similarly affected. The $h_{-}$ 
production cross section is that of the SM but reduced by a factor of 
$\cos^2(\vartheta-\theta)$, while the production cross section of 
$h_{+}$ is that of a SM Higgs \emph{of mass} $m_{+}$ reduced by a factor 
of $\sin^2(\vartheta-\theta)$.

\subsection{Twin Sector Higgs Decays}

In this section we provide expressions for the widths of some of the 
important decay modes of the twin sector Higgs. The couplings of the 125 
GeV boson are constrained to be quite SM like. To be consistent with 
these bounds we require $v_B\gtrsim 3v_\text{EW} $. Then the twin sector 
Higgs is heavy enough to decay to real pairs of visible sector top 
quarks and gauge bosons. In our analysis, we make use of the formulae 
gathered in \cite{Djouadi:2005gi}.

 The partial width for twin sector Higgs decays into fermions is given by
 \begin{equation}
\Gamma(h_{+}\to f_{A,B}f_{A,B})=\frac{N_c m_{+}Y_{f_{A,B}}^2}{16\pi}\left(1-4\frac{m_{f_{A,B}}^2}{m_{+}^2} \right)^{3/2}\left\{ \begin{array}{cc}
\sin^2(\vartheta-\theta) & f_A\\
\cos^2(\vartheta-\theta) & f_B
\end{array}\right..
 \end{equation}
 The corresponding expression for decays into on-shell pairs of $W$ and 
$Z$ gauge bosons is given by
 \begin{equation}
\Gamma(h_{+}\to V_{A,B} V_{A,B})=\frac{m_{+}^3\delta_V}{32\pi}\sqrt{1-4\frac{m_V^2}{m_{+}^2}}\left(1-4\frac{m_V^2}{m_{+}^2}+12\frac{m_V^4}{m_{+}^4} \right)\left\{ \begin{array}{cc}
\displaystyle\frac{\sin^2(\vartheta-\theta)}{v_A^2} & V_A\\[3mm]
\displaystyle\frac{\cos^2(\vartheta-\theta)}{v_B^2} & V_B
\end{array}\right. \;,
\label{HVV}
 \end{equation} 
 where $\delta_W=2$ and $\delta_Z=1$. If the twin sector Higgs is light, 
then decays to on-shell twin sectors gauge bosons may be kinematically 
forbidden. In this case, one of the gauge bosons can be off shell, with 
the virtual particle decaying to lighter fermions, $h_{+}\to VV^\ast\to 
Vff$. In the limit that the final state fermions are massless the 
corresponding partial widths are given by
 \begin{equation}
\Gamma(h_{+}\to V_B V_B^\ast)=\frac{3m_{+}m_V^4\delta'_V}{32\pi^3 v_B^4}R_T\left( \frac{m_V^2}{m_{+}^2}\right)\cos^2(\vartheta-\theta),
 \end{equation}
 where $\delta'_W=1$ and 
 \begin{equation}
 \delta'_Z=\frac{7}{12}-\frac{10}{9}\sin^2\theta_W+\frac{40}{9}\sin^4\theta_W \;.
 \end{equation}
 Here
 \begin{equation}
R_T(x)=\frac{3(1-8 x+20 x^2)}{\sqrt{4x-1}}\arccos\left( \frac{3x-1}{2x^{3/2}}\right)-\frac{1-x}{2x}(2-13x+47x^2)-\frac32(1-6x+4x^2)\ln x.
 \end{equation}

 We now turn to twin sector Higgs decays into light Higgs bosons. 
Expanding out the Higgs potential, Eq. \eqref{e.nonlinpot}, we obtain a 
contribution to the cubic $h_{+}h_{-}^2$ coupling,
 \begin{equation}
\sigma^3\sqrt{2}f\left[\lambda+\delta-\frac12\delta\sin^2(2\vartheta) \right]-\sigma^2h\frac{5\delta f}{2\sqrt{2}}\sin(4\vartheta) -\sigma h^2\sqrt{2}\delta f\left[1-3\sin^2(2\vartheta) \right]+h^3\frac{\delta f}{\sqrt{2}}\sin(4\vartheta).
 \end{equation}
 This leads to 
 \begin{align}
g_{h_{+}h_{-}h_{-}}\equiv&\frac{2f}{\sqrt{2}}\left\{3(\lambda+\delta)\sin\theta\sin(2\theta)-\frac{\delta}{8}\left[ \cos\theta-9\cos(3\theta)+2\cos(\theta-4\vartheta)\right.\right.\nonumber\\
&\left.\phantom{\frac{2f}{\sqrt{2}}}\left.+\cos(\theta+4\vartheta)+21\cos(3\theta-4\vartheta)\right]\right\}.
 \end{align}
 There is also a contribution to the decay width from the kinetic term,
 \begin{equation}
\frac{1}{\sqrt{2}f}\sigma\partial_\mu h\partial^\mu h.
 \end{equation}
 Then the partial width of $h_{+}$ into $h_{-}$ pairs is given by
 \begin{align}
\Gamma(h_{+}\to h_{-}h_{-})=\frac{1}{32\pi m_{+}}\sqrt{1-4\frac{m_{-}^2}{m_{+}^2}}&\left[ g_{h_{+}h_{-}h_{-}}+\frac{2}{\sqrt{2}f}\cos\theta(1+\sin^2\theta)\left(\frac{m_{+}^2}{2}-m_{-}^2 \right)\right.\nonumber\\
&\left.+\frac{4m_{-}^2}{\sqrt{2}f}\cos\theta\sin^2\theta \right]^2.
 \end{align}

\bibliography{TwinHiggsCancellationbib}

\begin{thebibliography}{78}%
\makeatletter
\providecommand \@ifxundefined [1]{%
 \@ifx{#1\undefined}
}%
\providecommand \@ifnum [1]{%
 \ifnum #1\expandafter \@firstoftwo
 \else \expandafter \@secondoftwo
 \fi
}%
\providecommand \@ifx [1]{%
 \ifx #1\expandafter \@firstoftwo
 \else \expandafter \@secondoftwo
 \fi
}%
\providecommand \natexlab [1]{#1}%
\providecommand \enquote  [1]{``#1''}%
\providecommand \bibnamefont  [1]{#1}%
\providecommand \bibfnamefont [1]{#1}%
\providecommand \citenamefont [1]{#1}%
\providecommand \href@noop [0]{\@secondoftwo}%
\providecommand \href [0]{\begingroup \@sanitize@url \@href}%
\providecommand \@href[1]{\@@startlink{#1}\@@href}%
\providecommand \@@href[1]{\endgroup#1\@@endlink}%
\providecommand \@sanitize@url [0]{\catcode `\\12\catcode `\$12\catcode
  `\&12\catcode `\#12\catcode `\^12\catcode `\_12\catcode `\%12\relax}%
\providecommand \@@startlink[1]{}%
\providecommand \@@endlink[0]{}%
\providecommand \url  [0]{\begingroup\@sanitize@url \@url }%
\providecommand \@url [1]{\endgroup\@href {#1}{\urlprefix }}%
\providecommand \urlprefix  [0]{URL }%
\providecommand \Eprint [0]{\href }%
\providecommand \doibase [0]{http://dx.doi.org/}%
\providecommand \selectlanguage [0]{\@gobble}%
\providecommand \bibinfo  [0]{\@secondoftwo}%
\providecommand \bibfield  [0]{\@secondoftwo}%
\providecommand \translation [1]{[#1]}%
\providecommand \BibitemOpen [0]{}%
\providecommand \bibitemStop [0]{}%
\providecommand \bibitemNoStop [0]{.\EOS\space}%
\providecommand \EOS [0]{\spacefactor3000\relax}%
\providecommand \BibitemShut  [1]{\csname bibitem#1\endcsname}%
\let\auto@bib@innerbib\@empty
\bibitem [{\citenamefont {Aad}\ \emph {et~al.}(2012)\citenamefont {Aad} \emph
  {et~al.}}]{Aad:2012tfa}%
  \BibitemOpen
  \bibfield  {author} {\bibinfo {author} {\bibfnamefont {G.}~\bibnamefont
  {Aad}} \emph {et~al.} (\bibinfo {collaboration} {ATLAS}),\ }\href {\doibase
  10.1016/j.physletb.2012.08.020} {\bibfield  {journal} {\bibinfo  {journal}
  {Phys.Lett.}\ }\textbf {\bibinfo {volume} {B716}},\ \bibinfo {pages} {1}
  (\bibinfo {year} {2012})},\ \Eprint {http://arxiv.org/abs/1207.7214}
  {arXiv:1207.7214 [hep-ex]} \BibitemShut {NoStop}%
\bibitem [{\citenamefont {Chatrchyan}\ \emph {et~al.}(2012)\citenamefont
  {Chatrchyan} \emph {et~al.}}]{Chatrchyan:2012ufa}%
  \BibitemOpen
  \bibfield  {author} {\bibinfo {author} {\bibfnamefont {S.}~\bibnamefont
  {Chatrchyan}} \emph {et~al.} (\bibinfo {collaboration} {CMS}),\ }\href
  {\doibase 10.1016/j.physletb.2012.08.021} {\bibfield  {journal} {\bibinfo
  {journal} {Phys.Lett.}\ }\textbf {\bibinfo {volume} {B716}},\ \bibinfo
  {pages} {30} (\bibinfo {year} {2012})},\ \Eprint
  {http://arxiv.org/abs/1207.7235} {arXiv:1207.7235 [hep-ex]} \BibitemShut
  {NoStop}%
\bibitem [{\citenamefont {Fayet}(1977)}]{Fayet:1977yc}%
  \BibitemOpen
  \bibfield  {author} {\bibinfo {author} {\bibfnamefont {P.}~\bibnamefont
  {Fayet}},\ }\href {\doibase 10.1016/0370-2693(77)90852-8} {\bibfield
  {journal} {\bibinfo  {journal} {Phys. Lett.}\ }\textbf {\bibinfo {volume}
  {69B}},\ \bibinfo {pages} {489} (\bibinfo {year} {1977})}\BibitemShut
  {NoStop}%
\bibitem [{\citenamefont {Dimopoulos}\ and\ \citenamefont
  {Georgi}(1981)}]{Dimopoulos:1981zb}%
  \BibitemOpen
  \bibfield  {author} {\bibinfo {author} {\bibfnamefont {S.}~\bibnamefont
  {Dimopoulos}}\ and\ \bibinfo {author} {\bibfnamefont {H.}~\bibnamefont
  {Georgi}},\ }\href {\doibase 10.1016/0550-3213(81)90522-8} {\bibfield
  {journal} {\bibinfo  {journal} {Nucl. Phys.}\ }\textbf {\bibinfo {volume}
  {B193}},\ \bibinfo {pages} {150} (\bibinfo {year} {1981})}\BibitemShut
  {NoStop}%
\bibitem [{\citenamefont {Arkani-Hamed}\ \emph {et~al.}(2001)\citenamefont
  {Arkani-Hamed}, \citenamefont {Cohen},\ and\ \citenamefont
  {Georgi}}]{ArkaniHamed:2001nc}%
  \BibitemOpen
  \bibfield  {author} {\bibinfo {author} {\bibfnamefont {N.}~\bibnamefont
  {Arkani-Hamed}}, \bibinfo {author} {\bibfnamefont {A.~G.}\ \bibnamefont
  {Cohen}}, \ and\ \bibinfo {author} {\bibfnamefont {H.}~\bibnamefont
  {Georgi}},\ }\href {\doibase 10.1016/S0370-2693(01)00741-9} {\bibfield
  {journal} {\bibinfo  {journal} {Phys. Lett.}\ }\textbf {\bibinfo {volume}
  {B513}},\ \bibinfo {pages} {232} (\bibinfo {year} {2001})},\ \Eprint
  {http://arxiv.org/abs/hep-ph/0105239} {arXiv:hep-ph/0105239 [hep-ph]}
  \BibitemShut {NoStop}%
\bibitem [{\citenamefont {Chacko}\ \emph
  {et~al.}(2006{\natexlab{a}})\citenamefont {Chacko}, \citenamefont {Goh},\
  and\ \citenamefont {Harnik}}]{Chacko:2005pe}%
  \BibitemOpen
  \bibfield  {author} {\bibinfo {author} {\bibfnamefont {Z.}~\bibnamefont
  {Chacko}}, \bibinfo {author} {\bibfnamefont {H.-S.}\ \bibnamefont {Goh}}, \
  and\ \bibinfo {author} {\bibfnamefont {R.}~\bibnamefont {Harnik}},\ }\href
  {\doibase 10.1103/PhysRevLett.96.231802} {\bibfield  {journal} {\bibinfo
  {journal} {Phys. Rev. Lett.}\ }\textbf {\bibinfo {volume} {96}},\ \bibinfo
  {pages} {231802} (\bibinfo {year} {2006}{\natexlab{a}})},\ \Eprint
  {http://arxiv.org/abs/hep-ph/0506256} {arXiv:hep-ph/0506256 [hep-ph]}
  \BibitemShut {NoStop}%
\bibitem [{\citenamefont {Barbieri}\ \emph {et~al.}(2005)\citenamefont
  {Barbieri}, \citenamefont {Gregoire},\ and\ \citenamefont
  {Hall}}]{Barbieri:2005ri}%
  \BibitemOpen
  \bibfield  {author} {\bibinfo {author} {\bibfnamefont {R.}~\bibnamefont
  {Barbieri}}, \bibinfo {author} {\bibfnamefont {T.}~\bibnamefont {Gregoire}},
  \ and\ \bibinfo {author} {\bibfnamefont {L.~J.}\ \bibnamefont {Hall}},\
  }\href@noop {} {\  (\bibinfo {year} {2005})},\ \Eprint
  {http://arxiv.org/abs/hep-ph/0509242} {arXiv:hep-ph/0509242 [hep-ph]}
  \BibitemShut {NoStop}%
\bibitem [{\citenamefont {Chacko}\ \emph
  {et~al.}(2006{\natexlab{b}})\citenamefont {Chacko}, \citenamefont {Nomura},
  \citenamefont {Papucci},\ and\ \citenamefont {Perez}}]{Chacko:2005vw}%
  \BibitemOpen
  \bibfield  {author} {\bibinfo {author} {\bibfnamefont {Z.}~\bibnamefont
  {Chacko}}, \bibinfo {author} {\bibfnamefont {Y.}~\bibnamefont {Nomura}},
  \bibinfo {author} {\bibfnamefont {M.}~\bibnamefont {Papucci}}, \ and\
  \bibinfo {author} {\bibfnamefont {G.}~\bibnamefont {Perez}},\ }\href
  {\doibase 10.1088/1126-6708/2006/01/126} {\bibfield  {journal} {\bibinfo
  {journal} {JHEP}\ }\textbf {\bibinfo {volume} {01}},\ \bibinfo {pages} {126}
  (\bibinfo {year} {2006}{\natexlab{b}})},\ \Eprint
  {http://arxiv.org/abs/hep-ph/0510273} {arXiv:hep-ph/0510273 [hep-ph]}
  \BibitemShut {NoStop}%
\bibitem [{\citenamefont {Burdman}\ \emph {et~al.}(2007)\citenamefont
  {Burdman}, \citenamefont {Chacko}, \citenamefont {Goh},\ and\ \citenamefont
  {Harnik}}]{Burdman:2006tz}%
  \BibitemOpen
  \bibfield  {author} {\bibinfo {author} {\bibfnamefont {G.}~\bibnamefont
  {Burdman}}, \bibinfo {author} {\bibfnamefont {Z.}~\bibnamefont {Chacko}},
  \bibinfo {author} {\bibfnamefont {H.-S.}\ \bibnamefont {Goh}}, \ and\
  \bibinfo {author} {\bibfnamefont {R.}~\bibnamefont {Harnik}},\ }\href
  {\doibase 10.1088/1126-6708/2007/02/009} {\bibfield  {journal} {\bibinfo
  {journal} {JHEP}\ }\textbf {\bibinfo {volume} {02}},\ \bibinfo {pages} {009}
  (\bibinfo {year} {2007})},\ \Eprint {http://arxiv.org/abs/hep-ph/0609152}
  {arXiv:hep-ph/0609152 [hep-ph]} \BibitemShut {NoStop}%
\bibitem [{\citenamefont {Cai}\ \emph {et~al.}(2009)\citenamefont {Cai},
  \citenamefont {Cheng},\ and\ \citenamefont {Terning}}]{Cai:2008au}%
  \BibitemOpen
  \bibfield  {author} {\bibinfo {author} {\bibfnamefont {H.}~\bibnamefont
  {Cai}}, \bibinfo {author} {\bibfnamefont {H.-C.}\ \bibnamefont {Cheng}}, \
  and\ \bibinfo {author} {\bibfnamefont {J.}~\bibnamefont {Terning}},\ }\href
  {\doibase 10.1088/1126-6708/2009/05/045} {\bibfield  {journal} {\bibinfo
  {journal} {JHEP}\ }\textbf {\bibinfo {volume} {05}},\ \bibinfo {pages} {045}
  (\bibinfo {year} {2009})},\ \Eprint {http://arxiv.org/abs/0812.0843}
  {arXiv:0812.0843 [hep-ph]} \BibitemShut {NoStop}%
\bibitem [{\citenamefont {Poland}\ and\ \citenamefont
  {Thaler}(2008)}]{Poland:2008ev}%
  \BibitemOpen
  \bibfield  {author} {\bibinfo {author} {\bibfnamefont {D.}~\bibnamefont
  {Poland}}\ and\ \bibinfo {author} {\bibfnamefont {J.}~\bibnamefont
  {Thaler}},\ }\href {\doibase 10.1088/1126-6708/2008/11/083} {\bibfield
  {journal} {\bibinfo  {journal} {JHEP}\ }\textbf {\bibinfo {volume} {11}},\
  \bibinfo {pages} {083} (\bibinfo {year} {2008})},\ \Eprint
  {http://arxiv.org/abs/0808.1290} {arXiv:0808.1290 [hep-ph]} \BibitemShut
  {NoStop}%
\bibitem [{\citenamefont {Batell}\ and\ \citenamefont
  {McCullough}(2015)}]{Batell:2015aha}%
  \BibitemOpen
  \bibfield  {author} {\bibinfo {author} {\bibfnamefont {B.}~\bibnamefont
  {Batell}}\ and\ \bibinfo {author} {\bibfnamefont {M.}~\bibnamefont
  {McCullough}},\ }\href {\doibase 10.1103/PhysRevD.92.073018} {\bibfield
  {journal} {\bibinfo  {journal} {Phys. Rev.}\ }\textbf {\bibinfo {volume}
  {D92}},\ \bibinfo {pages} {073018} (\bibinfo {year} {2015})},\ \Eprint
  {http://arxiv.org/abs/1504.04016} {arXiv:1504.04016 [hep-ph]} \BibitemShut
  {NoStop}%
\bibitem [{\citenamefont {Serra}\ and\ \citenamefont
  {Torre}(2017)}]{Serra:2017poj}%
  \BibitemOpen
  \bibfield  {author} {\bibinfo {author} {\bibfnamefont {J.}~\bibnamefont
  {Serra}}\ and\ \bibinfo {author} {\bibfnamefont {R.}~\bibnamefont {Torre}},\
  }\href@noop {} {\  (\bibinfo {year} {2017})},\ \Eprint
  {http://arxiv.org/abs/1709.05399} {arXiv:1709.05399 [hep-ph]} \BibitemShut
  {NoStop}%
\bibitem [{\citenamefont {Cs\'{a}ki}\ \emph {et~al.}(2017)\citenamefont
  {Cs\'{a}ki}, \citenamefont {Ma},\ and\ \citenamefont {Shu}}]{Csaki:2017jby}%
  \BibitemOpen
  \bibfield  {author} {\bibinfo {author} {\bibfnamefont {C.}~\bibnamefont
  {Cs\'{a}ki}}, \bibinfo {author} {\bibfnamefont {T.}~\bibnamefont {Ma}}, \
  and\ \bibinfo {author} {\bibfnamefont {J.}~\bibnamefont {Shu}},\ }\href@noop
  {} {\  (\bibinfo {year} {2017})},\ \Eprint {http://arxiv.org/abs/1709.08636}
  {arXiv:1709.08636 [hep-ph]} \BibitemShut {NoStop}%
\bibitem [{\citenamefont {Falkowski}\ \emph {et~al.}(2006)\citenamefont
  {Falkowski}, \citenamefont {Pokorski},\ and\ \citenamefont
  {Schmaltz}}]{Falkowski:2006qq}%
  \BibitemOpen
  \bibfield  {author} {\bibinfo {author} {\bibfnamefont {A.}~\bibnamefont
  {Falkowski}}, \bibinfo {author} {\bibfnamefont {S.}~\bibnamefont {Pokorski}},
  \ and\ \bibinfo {author} {\bibfnamefont {M.}~\bibnamefont {Schmaltz}},\
  }\href {\doibase 10.1103/PhysRevD.74.035003} {\bibfield  {journal} {\bibinfo
  {journal} {Phys. Rev.}\ }\textbf {\bibinfo {volume} {D74}},\ \bibinfo {pages}
  {035003} (\bibinfo {year} {2006})},\ \Eprint
  {http://arxiv.org/abs/hep-ph/0604066} {arXiv:hep-ph/0604066 [hep-ph]}
  \BibitemShut {NoStop}%
\bibitem [{\citenamefont {Chang}\ \emph {et~al.}(2007)\citenamefont {Chang},
  \citenamefont {Hall},\ and\ \citenamefont {Weiner}}]{chang:2006ra}%
  \BibitemOpen
  \bibfield  {author} {\bibinfo {author} {\bibfnamefont {S.}~\bibnamefont
  {Chang}}, \bibinfo {author} {\bibfnamefont {L.~J.}\ \bibnamefont {Hall}}, \
  and\ \bibinfo {author} {\bibfnamefont {N.}~\bibnamefont {Weiner}},\ }\href
  {\doibase 10.1103/PhysRevD.75.035009} {\bibfield  {journal} {\bibinfo
  {journal} {Phys. Rev.}\ }\textbf {\bibinfo {volume} {D75}},\ \bibinfo {pages}
  {035009} (\bibinfo {year} {2007})},\ \Eprint
  {http://arxiv.org/abs/hep-ph/0604076} {arXiv:hep-ph/0604076 [hep-ph]}
  \BibitemShut {NoStop}%
\bibitem [{\citenamefont {Craig}\ and\ \citenamefont
  {Howe}(2014)}]{Craig:2013fga}%
  \BibitemOpen
  \bibfield  {author} {\bibinfo {author} {\bibfnamefont {N.}~\bibnamefont
  {Craig}}\ and\ \bibinfo {author} {\bibfnamefont {K.}~\bibnamefont {Howe}},\
  }\href {\doibase 10.1007/JHEP03(2014)140} {\bibfield  {journal} {\bibinfo
  {journal} {JHEP}\ }\textbf {\bibinfo {volume} {03}},\ \bibinfo {pages} {140}
  (\bibinfo {year} {2014})},\ \Eprint {http://arxiv.org/abs/1312.1341}
  {arXiv:1312.1341 [hep-ph]} \BibitemShut {NoStop}%
\bibitem [{\citenamefont {Katz}\ \emph {et~al.}(2017)\citenamefont {Katz},
  \citenamefont {Mariotti}, \citenamefont {Pokorski}, \citenamefont
  {Redigolo},\ and\ \citenamefont {Ziegler}}]{Katz:2016wtw}%
  \BibitemOpen
  \bibfield  {author} {\bibinfo {author} {\bibfnamefont {A.}~\bibnamefont
  {Katz}}, \bibinfo {author} {\bibfnamefont {A.}~\bibnamefont {Mariotti}},
  \bibinfo {author} {\bibfnamefont {S.}~\bibnamefont {Pokorski}}, \bibinfo
  {author} {\bibfnamefont {D.}~\bibnamefont {Redigolo}}, \ and\ \bibinfo
  {author} {\bibfnamefont {R.}~\bibnamefont {Ziegler}},\ }\href {\doibase
  10.1007/JHEP01(2017)142} {\bibfield  {journal} {\bibinfo  {journal} {JHEP}\
  }\textbf {\bibinfo {volume} {01}},\ \bibinfo {pages} {142} (\bibinfo {year}
  {2017})},\ \Eprint {http://arxiv.org/abs/1611.08615} {arXiv:1611.08615
  [hep-ph]} \BibitemShut {NoStop}%
\bibitem [{\citenamefont {Badziak}\ and\ \citenamefont
  {Harigaya}(2017{\natexlab{a}})}]{Badziak:2017syq}%
  \BibitemOpen
  \bibfield  {author} {\bibinfo {author} {\bibfnamefont {M.}~\bibnamefont
  {Badziak}}\ and\ \bibinfo {author} {\bibfnamefont {K.}~\bibnamefont
  {Harigaya}},\ }\href {\doibase 10.1007/JHEP06(2017)065} {\bibfield  {journal}
  {\bibinfo  {journal} {JHEP}\ }\textbf {\bibinfo {volume} {06}},\ \bibinfo
  {pages} {065} (\bibinfo {year} {2017}{\natexlab{a}})},\ \Eprint
  {http://arxiv.org/abs/1703.02122} {arXiv:1703.02122 [hep-ph]} \BibitemShut
  {NoStop}%
\bibitem [{\citenamefont {Badziak}\ and\ \citenamefont
  {Harigaya}(2017{\natexlab{b}})}]{Badziak:2017kjk}%
  \BibitemOpen
  \bibfield  {author} {\bibinfo {author} {\bibfnamefont {M.}~\bibnamefont
  {Badziak}}\ and\ \bibinfo {author} {\bibfnamefont {K.}~\bibnamefont
  {Harigaya}},\ }\href {\doibase 10.1007/JHEP10(2017)109} {\bibfield  {journal}
  {\bibinfo  {journal} {JHEP}\ }\textbf {\bibinfo {volume} {10}},\ \bibinfo
  {pages} {109} (\bibinfo {year} {2017}{\natexlab{b}})},\ \Eprint
  {http://arxiv.org/abs/1707.09071} {arXiv:1707.09071 [hep-ph]} \BibitemShut
  {NoStop}%
\bibitem [{\citenamefont {Geller}\ and\ \citenamefont
  {Telem}(2015)}]{Geller:2014kta}%
  \BibitemOpen
  \bibfield  {author} {\bibinfo {author} {\bibfnamefont {M.}~\bibnamefont
  {Geller}}\ and\ \bibinfo {author} {\bibfnamefont {O.}~\bibnamefont {Telem}},\
  }\href {\doibase 10.1103/PhysRevLett.114.191801} {\bibfield  {journal}
  {\bibinfo  {journal} {Phys. Rev. Lett.}\ }\textbf {\bibinfo {volume} {114}},\
  \bibinfo {pages} {191801} (\bibinfo {year} {2015})},\ \Eprint
  {http://arxiv.org/abs/1411.2974} {arXiv:1411.2974 [hep-ph]} \BibitemShut
  {NoStop}%
\bibitem [{\citenamefont {Barbieri}\ \emph {et~al.}(2015)\citenamefont
  {Barbieri}, \citenamefont {Greco}, \citenamefont {Rattazzi},\ and\
  \citenamefont {Wulzer}}]{Barbieri:2015lqa}%
  \BibitemOpen
  \bibfield  {author} {\bibinfo {author} {\bibfnamefont {R.}~\bibnamefont
  {Barbieri}}, \bibinfo {author} {\bibfnamefont {D.}~\bibnamefont {Greco}},
  \bibinfo {author} {\bibfnamefont {R.}~\bibnamefont {Rattazzi}}, \ and\
  \bibinfo {author} {\bibfnamefont {A.}~\bibnamefont {Wulzer}},\ }\href
  {\doibase 10.1007/JHEP08(2015)161} {\bibfield  {journal} {\bibinfo  {journal}
  {JHEP}\ }\textbf {\bibinfo {volume} {08}},\ \bibinfo {pages} {161} (\bibinfo
  {year} {2015})},\ \Eprint {http://arxiv.org/abs/1501.07803} {arXiv:1501.07803
  [hep-ph]} \BibitemShut {NoStop}%
\bibitem [{\citenamefont {Low}\ \emph {et~al.}(2015)\citenamefont {Low},
  \citenamefont {Tesi},\ and\ \citenamefont {Wang}}]{Low:2015nqa}%
  \BibitemOpen
  \bibfield  {author} {\bibinfo {author} {\bibfnamefont {M.}~\bibnamefont
  {Low}}, \bibinfo {author} {\bibfnamefont {A.}~\bibnamefont {Tesi}}, \ and\
  \bibinfo {author} {\bibfnamefont {L.-T.}\ \bibnamefont {Wang}},\ }\href
  {\doibase 10.1103/PhysRevD.91.095012} {\bibfield  {journal} {\bibinfo
  {journal} {Phys. Rev.}\ }\textbf {\bibinfo {volume} {D91}},\ \bibinfo {pages}
  {095012} (\bibinfo {year} {2015})},\ \Eprint
  {http://arxiv.org/abs/1501.07890} {arXiv:1501.07890 [hep-ph]} \BibitemShut
  {NoStop}%
\bibitem [{\citenamefont {Cheng}\ \emph {et~al.}(2016)\citenamefont {Cheng},
  \citenamefont {Jung}, \citenamefont {Salvioni},\ and\ \citenamefont
  {Tsai}}]{Cheng:2015buv}%
  \BibitemOpen
  \bibfield  {author} {\bibinfo {author} {\bibfnamefont {H.-C.}\ \bibnamefont
  {Cheng}}, \bibinfo {author} {\bibfnamefont {S.}~\bibnamefont {Jung}},
  \bibinfo {author} {\bibfnamefont {E.}~\bibnamefont {Salvioni}}, \ and\
  \bibinfo {author} {\bibfnamefont {Y.}~\bibnamefont {Tsai}},\ }\href {\doibase
  10.1007/JHEP03(2016)074} {\bibfield  {journal} {\bibinfo  {journal} {JHEP}\
  }\textbf {\bibinfo {volume} {03}},\ \bibinfo {pages} {074} (\bibinfo {year}
  {2016})},\ \Eprint {http://arxiv.org/abs/1512.02647} {arXiv:1512.02647
  [hep-ph]} \BibitemShut {NoStop}%
\bibitem [{\citenamefont {Cheng}\ \emph {et~al.}(2017)\citenamefont {Cheng},
  \citenamefont {Salvioni},\ and\ \citenamefont {Tsai}}]{Cheng:2016uqk}%
  \BibitemOpen
  \bibfield  {author} {\bibinfo {author} {\bibfnamefont {H.-C.}\ \bibnamefont
  {Cheng}}, \bibinfo {author} {\bibfnamefont {E.}~\bibnamefont {Salvioni}}, \
  and\ \bibinfo {author} {\bibfnamefont {Y.}~\bibnamefont {Tsai}},\ }\href
  {\doibase 10.1103/PhysRevD.95.115035} {\bibfield  {journal} {\bibinfo
  {journal} {Phys. Rev.}\ }\textbf {\bibinfo {volume} {D95}},\ \bibinfo {pages}
  {115035} (\bibinfo {year} {2017})},\ \Eprint
  {http://arxiv.org/abs/1612.03176} {arXiv:1612.03176 [hep-ph]} \BibitemShut
  {NoStop}%
\bibitem [{\citenamefont {Contino}\ \emph {et~al.}(2017)\citenamefont
  {Contino}, \citenamefont {Greco}, \citenamefont {Mahbubani}, \citenamefont
  {Rattazzi},\ and\ \citenamefont {Torre}}]{Contino:2017moj}%
  \BibitemOpen
  \bibfield  {author} {\bibinfo {author} {\bibfnamefont {R.}~\bibnamefont
  {Contino}}, \bibinfo {author} {\bibfnamefont {D.}~\bibnamefont {Greco}},
  \bibinfo {author} {\bibfnamefont {R.}~\bibnamefont {Mahbubani}}, \bibinfo
  {author} {\bibfnamefont {R.}~\bibnamefont {Rattazzi}}, \ and\ \bibinfo
  {author} {\bibfnamefont {R.}~\bibnamefont {Torre}},\ }\href@noop {} {\
  (\bibinfo {year} {2017})},\ \Eprint {http://arxiv.org/abs/1702.00797}
  {arXiv:1702.00797 [hep-ph]} \BibitemShut {NoStop}%
\bibitem [{\citenamefont {Csaki}\ \emph {et~al.}(2016)\citenamefont {Csaki},
  \citenamefont {Geller}, \citenamefont {Telem},\ and\ \citenamefont
  {Weiler}}]{Csaki:2015gfd}%
  \BibitemOpen
  \bibfield  {author} {\bibinfo {author} {\bibfnamefont {C.}~\bibnamefont
  {Csaki}}, \bibinfo {author} {\bibfnamefont {M.}~\bibnamefont {Geller}},
  \bibinfo {author} {\bibfnamefont {O.}~\bibnamefont {Telem}}, \ and\ \bibinfo
  {author} {\bibfnamefont {A.}~\bibnamefont {Weiler}},\ }\href {\doibase
  10.1007/JHEP09(2016)146} {\bibfield  {journal} {\bibinfo  {journal} {JHEP}\
  }\textbf {\bibinfo {volume} {09}},\ \bibinfo {pages} {146} (\bibinfo {year}
  {2016})},\ \Eprint {http://arxiv.org/abs/1512.03427} {arXiv:1512.03427
  [hep-ph]} \BibitemShut {NoStop}%
\bibitem [{\citenamefont {Beauchesne}\ \emph {et~al.}(2016)\citenamefont
  {Beauchesne}, \citenamefont {Earl},\ and\ \citenamefont
  {Gr\'{e}goire}}]{Beauchesne:2015lva}%
  \BibitemOpen
  \bibfield  {author} {\bibinfo {author} {\bibfnamefont {H.}~\bibnamefont
  {Beauchesne}}, \bibinfo {author} {\bibfnamefont {K.}~\bibnamefont {Earl}}, \
  and\ \bibinfo {author} {\bibfnamefont {T.}~\bibnamefont {Gr\'{e}goire}},\
  }\href {\doibase 10.1007/JHEP01(2016)130} {\bibfield  {journal} {\bibinfo
  {journal} {JHEP}\ }\textbf {\bibinfo {volume} {01}},\ \bibinfo {pages} {130}
  (\bibinfo {year} {2016})},\ \Eprint {http://arxiv.org/abs/1510.06069}
  {arXiv:1510.06069 [hep-ph]} \BibitemShut {NoStop}%
\bibitem [{\citenamefont {Harnik}\ \emph {et~al.}(2017)\citenamefont {Harnik},
  \citenamefont {Howe},\ and\ \citenamefont {Kearney}}]{Harnik:2016koz}%
  \BibitemOpen
  \bibfield  {author} {\bibinfo {author} {\bibfnamefont {R.}~\bibnamefont
  {Harnik}}, \bibinfo {author} {\bibfnamefont {K.}~\bibnamefont {Howe}}, \ and\
  \bibinfo {author} {\bibfnamefont {J.}~\bibnamefont {Kearney}},\ }\href
  {\doibase 10.1007/JHEP03(2017)111} {\bibfield  {journal} {\bibinfo  {journal}
  {JHEP}\ }\textbf {\bibinfo {volume} {03}},\ \bibinfo {pages} {111} (\bibinfo
  {year} {2017})},\ \Eprint {http://arxiv.org/abs/1603.03772} {arXiv:1603.03772
  [hep-ph]} \BibitemShut {NoStop}%
\bibitem [{\citenamefont {Yu}(2016{\natexlab{a}})}]{Yu:2016bku}%
  \BibitemOpen
  \bibfield  {author} {\bibinfo {author} {\bibfnamefont {J.-H.}\ \bibnamefont
  {Yu}},\ }\href {\doibase 10.1103/PhysRevD.94.111704} {\bibfield  {journal}
  {\bibinfo  {journal} {Phys. Rev.}\ }\textbf {\bibinfo {volume} {D94}},\
  \bibinfo {pages} {111704} (\bibinfo {year} {2016}{\natexlab{a}})},\ \Eprint
  {http://arxiv.org/abs/1608.01314} {arXiv:1608.01314 [hep-ph]} \BibitemShut
  {NoStop}%
\bibitem [{\citenamefont {Yu}(2016{\natexlab{b}})}]{Yu:2016swa}%
  \BibitemOpen
  \bibfield  {author} {\bibinfo {author} {\bibfnamefont {J.-H.}\ \bibnamefont
  {Yu}},\ }\href {\doibase 10.1007/JHEP12(2016)143} {\bibfield  {journal}
  {\bibinfo  {journal} {JHEP}\ }\textbf {\bibinfo {volume} {12}},\ \bibinfo
  {pages} {143} (\bibinfo {year} {2016}{\natexlab{b}})},\ \Eprint
  {http://arxiv.org/abs/1608.05713} {arXiv:1608.05713 [hep-ph]} \BibitemShut
  {NoStop}%
\bibitem [{\citenamefont {Yu}(2017)}]{Yu:2016cdr}%
  \BibitemOpen
  \bibfield  {author} {\bibinfo {author} {\bibfnamefont {J.-H.}\ \bibnamefont
  {Yu}},\ }\href@noop {} {\bibfield  {journal} {\bibinfo  {journal} {Phys.
  Rev.}\ }\textbf {\bibinfo {volume} {D95}},\ \bibinfo {pages} {095028}
  (\bibinfo {year} {2017})},\ \Eprint {http://arxiv.org/abs/1612.09300}
  {arXiv:1612.09300 [hep-ph]} \BibitemShut {NoStop}%
\bibitem [{\citenamefont {Craig}\ \emph
  {et~al.}(2015{\natexlab{a}})\citenamefont {Craig}, \citenamefont {Knapen},\
  and\ \citenamefont {Longhi}}]{Craig:2014aea}%
  \BibitemOpen
  \bibfield  {author} {\bibinfo {author} {\bibfnamefont {N.}~\bibnamefont
  {Craig}}, \bibinfo {author} {\bibfnamefont {S.}~\bibnamefont {Knapen}}, \
  and\ \bibinfo {author} {\bibfnamefont {P.}~\bibnamefont {Longhi}},\ }\href
  {\doibase 10.1103/PhysRevLett.114.061803} {\bibfield  {journal} {\bibinfo
  {journal} {Phys. Rev. Lett.}\ }\textbf {\bibinfo {volume} {114}},\ \bibinfo
  {pages} {061803} (\bibinfo {year} {2015}{\natexlab{a}})},\ \Eprint
  {http://arxiv.org/abs/1410.6808} {arXiv:1410.6808 [hep-ph]} \BibitemShut
  {NoStop}%
\bibitem [{\citenamefont {Craig}\ \emph
  {et~al.}(2015{\natexlab{b}})\citenamefont {Craig}, \citenamefont {Knapen},\
  and\ \citenamefont {Longhi}}]{Craig:2014roa}%
  \BibitemOpen
  \bibfield  {author} {\bibinfo {author} {\bibfnamefont {N.}~\bibnamefont
  {Craig}}, \bibinfo {author} {\bibfnamefont {S.}~\bibnamefont {Knapen}}, \
  and\ \bibinfo {author} {\bibfnamefont {P.}~\bibnamefont {Longhi}},\ }\href
  {\doibase 10.1007/JHEP03(2015)106} {\bibfield  {journal} {\bibinfo  {journal}
  {JHEP}\ }\textbf {\bibinfo {volume} {03}},\ \bibinfo {pages} {106} (\bibinfo
  {year} {2015}{\natexlab{b}})},\ \Eprint {http://arxiv.org/abs/1411.7393}
  {arXiv:1411.7393 [hep-ph]} \BibitemShut {NoStop}%
\bibitem [{\citenamefont {Craig}\ \emph {et~al.}(2016)\citenamefont {Craig},
  \citenamefont {Knapen}, \citenamefont {Longhi},\ and\ \citenamefont
  {Strassler}}]{Craig:2016kue}%
  \BibitemOpen
  \bibfield  {author} {\bibinfo {author} {\bibfnamefont {N.}~\bibnamefont
  {Craig}}, \bibinfo {author} {\bibfnamefont {S.}~\bibnamefont {Knapen}},
  \bibinfo {author} {\bibfnamefont {P.}~\bibnamefont {Longhi}}, \ and\ \bibinfo
  {author} {\bibfnamefont {M.}~\bibnamefont {Strassler}},\ }\href {\doibase
  10.1007/JHEP07(2016)002} {\bibfield  {journal} {\bibinfo  {journal} {JHEP}\
  }\textbf {\bibinfo {volume} {07}},\ \bibinfo {pages} {002} (\bibinfo {year}
  {2016})},\ \Eprint {http://arxiv.org/abs/1601.07181} {arXiv:1601.07181
  [hep-ph]} \BibitemShut {NoStop}%
\bibitem [{\citenamefont {Thrasher}(2017)}]{Thrasher:2017rpa}%
  \BibitemOpen
  \bibfield  {author} {\bibinfo {author} {\bibfnamefont {K.}~\bibnamefont
  {Thrasher}},\ }\href@noop {} {\  (\bibinfo {year} {2017})},\ \Eprint
  {http://arxiv.org/abs/1705.01472} {arXiv:1705.01472 [hep-ph]} \BibitemShut
  {NoStop}%
\bibitem [{\citenamefont {Chacko}\ \emph {et~al.}(2017)\citenamefont {Chacko},
  \citenamefont {Craig}, \citenamefont {Fox},\ and\ \citenamefont
  {Harnik}}]{Chacko:2016hvu}%
  \BibitemOpen
  \bibfield  {author} {\bibinfo {author} {\bibfnamefont {Z.}~\bibnamefont
  {Chacko}}, \bibinfo {author} {\bibfnamefont {N.}~\bibnamefont {Craig}},
  \bibinfo {author} {\bibfnamefont {P.~J.}\ \bibnamefont {Fox}}, \ and\
  \bibinfo {author} {\bibfnamefont {R.}~\bibnamefont {Harnik}},\ }\href
  {\doibase 10.1007/JHEP07(2017)023} {\bibfield  {journal} {\bibinfo  {journal}
  {JHEP}\ }\textbf {\bibinfo {volume} {07}},\ \bibinfo {pages} {023} (\bibinfo
  {year} {2017})},\ \Eprint {http://arxiv.org/abs/1611.07975} {arXiv:1611.07975
  [hep-ph]} \BibitemShut {NoStop}%
\bibitem [{\citenamefont {Craig}\ \emph {et~al.}(2017)\citenamefont {Craig},
  \citenamefont {Koren},\ and\ \citenamefont {Trott}}]{Craig:2016lyx}%
  \BibitemOpen
  \bibfield  {author} {\bibinfo {author} {\bibfnamefont {N.}~\bibnamefont
  {Craig}}, \bibinfo {author} {\bibfnamefont {S.}~\bibnamefont {Koren}}, \ and\
  \bibinfo {author} {\bibfnamefont {T.}~\bibnamefont {Trott}},\ }\href
  {\doibase 10.1007/JHEP05(2017)038} {\bibfield  {journal} {\bibinfo  {journal}
  {JHEP}\ }\textbf {\bibinfo {volume} {05}},\ \bibinfo {pages} {038} (\bibinfo
  {year} {2017})},\ \Eprint {http://arxiv.org/abs/1611.07977} {arXiv:1611.07977
  [hep-ph]} \BibitemShut {NoStop}%
\bibitem [{\citenamefont {Farina}(2015)}]{Farina:2015uea}%
  \BibitemOpen
  \bibfield  {author} {\bibinfo {author} {\bibfnamefont {M.}~\bibnamefont
  {Farina}},\ }\href {\doibase 10.1088/1475-7516/2015/11/017} {\bibfield
  {journal} {\bibinfo  {journal} {JCAP}\ }\textbf {\bibinfo {volume} {1511}},\
  \bibinfo {pages} {017} (\bibinfo {year} {2015})},\ \Eprint
  {http://arxiv.org/abs/1506.03520} {arXiv:1506.03520 [hep-ph]} \BibitemShut
  {NoStop}%
\bibitem [{\citenamefont {Barbieri}\ \emph {et~al.}(2016)\citenamefont
  {Barbieri}, \citenamefont {Hall},\ and\ \citenamefont
  {Harigaya}}]{Barbieri:2016zxn}%
  \BibitemOpen
  \bibfield  {author} {\bibinfo {author} {\bibfnamefont {R.}~\bibnamefont
  {Barbieri}}, \bibinfo {author} {\bibfnamefont {L.~J.}\ \bibnamefont {Hall}},
  \ and\ \bibinfo {author} {\bibfnamefont {K.}~\bibnamefont {Harigaya}},\
  }\href {\doibase 10.1007/JHEP11(2016)172} {\bibfield  {journal} {\bibinfo
  {journal} {JHEP}\ }\textbf {\bibinfo {volume} {11}},\ \bibinfo {pages} {172}
  (\bibinfo {year} {2016})},\ \Eprint {http://arxiv.org/abs/1609.05589}
  {arXiv:1609.05589 [hep-ph]} \BibitemShut {NoStop}%
\bibitem [{\citenamefont {Csaki}\ \emph {et~al.}(2017)\citenamefont {Csaki},
  \citenamefont {Kuflik},\ and\ \citenamefont {Lombardo}}]{Csaki:2017spo}%
  \BibitemOpen
  \bibfield  {author} {\bibinfo {author} {\bibfnamefont {C.}~\bibnamefont
  {Csaki}}, \bibinfo {author} {\bibfnamefont {E.}~\bibnamefont {Kuflik}}, \
  and\ \bibinfo {author} {\bibfnamefont {S.}~\bibnamefont {Lombardo}},\ }\href
  {\doibase 10.1103/PhysRevD.96.055013} {\bibfield  {journal} {\bibinfo
  {journal} {Phys. Rev.}\ }\textbf {\bibinfo {volume} {D96}},\ \bibinfo {pages}
  {055013} (\bibinfo {year} {2017})},\ \Eprint
  {http://arxiv.org/abs/1703.06884} {arXiv:1703.06884 [hep-ph]} \BibitemShut
  {NoStop}%
\bibitem [{\citenamefont {Barbieri}\ \emph {et~al.}(2017)\citenamefont
  {Barbieri}, \citenamefont {Hall},\ and\ \citenamefont
  {Harigaya}}]{Barbieri:2017opf}%
  \BibitemOpen
  \bibfield  {author} {\bibinfo {author} {\bibfnamefont {R.}~\bibnamefont
  {Barbieri}}, \bibinfo {author} {\bibfnamefont {L.~J.}\ \bibnamefont {Hall}},
  \ and\ \bibinfo {author} {\bibfnamefont {K.}~\bibnamefont {Harigaya}},\
  }\href {\doibase 10.1007/JHEP10(2017)015} {\bibfield  {journal} {\bibinfo
  {journal} {JHEP}\ }\textbf {\bibinfo {volume} {10}},\ \bibinfo {pages} {015}
  (\bibinfo {year} {2017})},\ \Eprint {http://arxiv.org/abs/1706.05548}
  {arXiv:1706.05548 [hep-ph]} \BibitemShut {NoStop}%
\bibitem [{\citenamefont {Farina}\ \emph {et~al.}(2016)\citenamefont {Farina},
  \citenamefont {Monteux},\ and\ \citenamefont {Shin}}]{Farina:2016ndq}%
  \BibitemOpen
  \bibfield  {author} {\bibinfo {author} {\bibfnamefont {M.}~\bibnamefont
  {Farina}}, \bibinfo {author} {\bibfnamefont {A.}~\bibnamefont {Monteux}}, \
  and\ \bibinfo {author} {\bibfnamefont {C.~S.}\ \bibnamefont {Shin}},\ }\href
  {\doibase 10.1103/PhysRevD.94.035017} {\bibfield  {journal} {\bibinfo
  {journal} {Phys. Rev.}\ }\textbf {\bibinfo {volume} {D94}},\ \bibinfo {pages}
  {035017} (\bibinfo {year} {2016})},\ \Eprint
  {http://arxiv.org/abs/1604.08211} {arXiv:1604.08211 [hep-ph]} \BibitemShut
  {NoStop}%
\bibitem [{\citenamefont {Craig}\ \emph
  {et~al.}(2015{\natexlab{c}})\citenamefont {Craig}, \citenamefont {Katz},
  \citenamefont {Strassler},\ and\ \citenamefont {Sundrum}}]{Craig:2015pha}%
  \BibitemOpen
  \bibfield  {author} {\bibinfo {author} {\bibfnamefont {N.}~\bibnamefont
  {Craig}}, \bibinfo {author} {\bibfnamefont {A.}~\bibnamefont {Katz}},
  \bibinfo {author} {\bibfnamefont {M.}~\bibnamefont {Strassler}}, \ and\
  \bibinfo {author} {\bibfnamefont {R.}~\bibnamefont {Sundrum}},\ }\href
  {\doibase 10.1007/JHEP07(2015)105} {\bibfield  {journal} {\bibinfo  {journal}
  {JHEP}\ }\textbf {\bibinfo {volume} {07}},\ \bibinfo {pages} {105} (\bibinfo
  {year} {2015}{\natexlab{c}})},\ \Eprint {http://arxiv.org/abs/1501.05310}
  {arXiv:1501.05310 [hep-ph]} \BibitemShut {NoStop}%
\bibitem [{\citenamefont {Curtin}\ and\ \citenamefont
  {Verhaaren}(2015)}]{Curtin:2015fna}%
  \BibitemOpen
  \bibfield  {author} {\bibinfo {author} {\bibfnamefont {D.}~\bibnamefont
  {Curtin}}\ and\ \bibinfo {author} {\bibfnamefont {C.~B.}\ \bibnamefont
  {Verhaaren}},\ }\href {\doibase 10.1007/JHEP12(2015)072} {\bibfield
  {journal} {\bibinfo  {journal} {JHEP}\ }\textbf {\bibinfo {volume} {12}},\
  \bibinfo {pages} {072} (\bibinfo {year} {2015})},\ \Eprint
  {http://arxiv.org/abs/1506.06141} {arXiv:1506.06141 [hep-ph]} \BibitemShut
  {NoStop}%
\bibitem [{\citenamefont {Csaki}\ \emph {et~al.}(2015)\citenamefont {Csaki},
  \citenamefont {Kuflik}, \citenamefont {Lombardo},\ and\ \citenamefont
  {Slone}}]{Csaki:2015fba}%
  \BibitemOpen
  \bibfield  {author} {\bibinfo {author} {\bibfnamefont {C.}~\bibnamefont
  {Csaki}}, \bibinfo {author} {\bibfnamefont {E.}~\bibnamefont {Kuflik}},
  \bibinfo {author} {\bibfnamefont {S.}~\bibnamefont {Lombardo}}, \ and\
  \bibinfo {author} {\bibfnamefont {O.}~\bibnamefont {Slone}},\ }\href
  {\doibase 10.1103/PhysRevD.92.073008} {\bibfield  {journal} {\bibinfo
  {journal} {Phys. Rev.}\ }\textbf {\bibinfo {volume} {D92}},\ \bibinfo {pages}
  {073008} (\bibinfo {year} {2015})},\ \Eprint
  {http://arxiv.org/abs/1508.01522} {arXiv:1508.01522 [hep-ph]} \BibitemShut
  {NoStop}%
\bibitem [{\citenamefont {Chou}\ \emph {et~al.}(2017)\citenamefont {Chou},
  \citenamefont {Curtin},\ and\ \citenamefont {Lubatti}}]{Chou:2016lxi}%
  \BibitemOpen
  \bibfield  {author} {\bibinfo {author} {\bibfnamefont {J.~P.}\ \bibnamefont
  {Chou}}, \bibinfo {author} {\bibfnamefont {D.}~\bibnamefont {Curtin}}, \ and\
  \bibinfo {author} {\bibfnamefont {H.~J.}\ \bibnamefont {Lubatti}},\ }\href
  {\doibase 10.1016/j.physletb.2017.01.043} {\bibfield  {journal} {\bibinfo
  {journal} {Phys. Lett.}\ }\textbf {\bibinfo {volume} {B767}},\ \bibinfo
  {pages} {29} (\bibinfo {year} {2017})},\ \Eprint
  {http://arxiv.org/abs/1606.06298} {arXiv:1606.06298 [hep-ph]} \BibitemShut
  {NoStop}%
\bibitem [{\citenamefont {Curtin}\ and\ \citenamefont
  {Peskin}(2017)}]{Curtin:2017izq}%
  \BibitemOpen
  \bibfield  {author} {\bibinfo {author} {\bibfnamefont {D.}~\bibnamefont
  {Curtin}}\ and\ \bibinfo {author} {\bibfnamefont {M.~E.}\ \bibnamefont
  {Peskin}},\ }\href@noop {} {\  (\bibinfo {year} {2017})},\ \Eprint
  {http://arxiv.org/abs/1705.06327} {arXiv:1705.06327 [hep-ph]} \BibitemShut
  {NoStop}%
\bibitem [{\citenamefont {Craig}\ and\ \citenamefont
  {Katz}(2015)}]{Craig:2015xla}%
  \BibitemOpen
  \bibfield  {author} {\bibinfo {author} {\bibfnamefont {N.}~\bibnamefont
  {Craig}}\ and\ \bibinfo {author} {\bibfnamefont {A.}~\bibnamefont {Katz}},\
  }\href {\doibase 10.1088/1475-7516/2015/10/054} {\bibfield  {journal}
  {\bibinfo  {journal} {JCAP}\ }\textbf {\bibinfo {volume} {1510}},\ \bibinfo
  {pages} {054} (\bibinfo {year} {2015})},\ \Eprint
  {http://arxiv.org/abs/1505.07113} {arXiv:1505.07113 [hep-ph]} \BibitemShut
  {NoStop}%
\bibitem [{\citenamefont {Garcia~Garcia}\ \emph
  {et~al.}(2015{\natexlab{a}})\citenamefont {Garcia~Garcia}, \citenamefont
  {Lasenby},\ and\ \citenamefont {March-Russell}}]{Garcia:2015loa}%
  \BibitemOpen
  \bibfield  {author} {\bibinfo {author} {\bibfnamefont {I.}~\bibnamefont
  {Garcia~Garcia}}, \bibinfo {author} {\bibfnamefont {R.}~\bibnamefont
  {Lasenby}}, \ and\ \bibinfo {author} {\bibfnamefont {J.}~\bibnamefont
  {March-Russell}},\ }\href {\doibase 10.1103/PhysRevD.92.055034} {\bibfield
  {journal} {\bibinfo  {journal} {Phys. Rev.}\ }\textbf {\bibinfo {volume}
  {D92}},\ \bibinfo {pages} {055034} (\bibinfo {year} {2015}{\natexlab{a}})},\
  \Eprint {http://arxiv.org/abs/1505.07109} {arXiv:1505.07109 [hep-ph]}
  \BibitemShut {NoStop}%
\bibitem [{\citenamefont {Garcia~Garcia}\ \emph
  {et~al.}(2015{\natexlab{b}})\citenamefont {Garcia~Garcia}, \citenamefont
  {Lasenby},\ and\ \citenamefont {March-Russell}}]{Garcia:2015toa}%
  \BibitemOpen
  \bibfield  {author} {\bibinfo {author} {\bibfnamefont {I.}~\bibnamefont
  {Garcia~Garcia}}, \bibinfo {author} {\bibfnamefont {R.}~\bibnamefont
  {Lasenby}}, \ and\ \bibinfo {author} {\bibfnamefont {J.}~\bibnamefont
  {March-Russell}},\ }\href {\doibase 10.1103/PhysRevLett.115.121801}
  {\bibfield  {journal} {\bibinfo  {journal} {Phys. Rev. Lett.}\ }\textbf
  {\bibinfo {volume} {115}},\ \bibinfo {pages} {121801} (\bibinfo {year}
  {2015}{\natexlab{b}})},\ \Eprint {http://arxiv.org/abs/1505.07410}
  {arXiv:1505.07410 [hep-ph]} \BibitemShut {NoStop}%
\bibitem [{\citenamefont {Freytsis}\ \emph {et~al.}(2016)\citenamefont
  {Freytsis}, \citenamefont {Knapen}, \citenamefont {Robinson},\ and\
  \citenamefont {Tsai}}]{Freytsis:2016dgf}%
  \BibitemOpen
  \bibfield  {author} {\bibinfo {author} {\bibfnamefont {M.}~\bibnamefont
  {Freytsis}}, \bibinfo {author} {\bibfnamefont {S.}~\bibnamefont {Knapen}},
  \bibinfo {author} {\bibfnamefont {D.~J.}\ \bibnamefont {Robinson}}, \ and\
  \bibinfo {author} {\bibfnamefont {Y.}~\bibnamefont {Tsai}},\ }\href {\doibase
  10.1007/JHEP05(2016)018} {\bibfield  {journal} {\bibinfo  {journal} {JHEP}\
  }\textbf {\bibinfo {volume} {05}},\ \bibinfo {pages} {018} (\bibinfo {year}
  {2016})},\ \Eprint {http://arxiv.org/abs/1601.07556} {arXiv:1601.07556
  [hep-ph]} \BibitemShut {NoStop}%
\bibitem [{\citenamefont {Prilepina}\ and\ \citenamefont
  {Tsai}(2017)}]{Prilepina:2016rlq}%
  \BibitemOpen
  \bibfield  {author} {\bibinfo {author} {\bibfnamefont {V.}~\bibnamefont
  {Prilepina}}\ and\ \bibinfo {author} {\bibfnamefont {Y.}~\bibnamefont
  {Tsai}},\ }\href {\doibase 10.1007/JHEP09(2017)033} {\bibfield  {journal}
  {\bibinfo  {journal} {JHEP}\ }\textbf {\bibinfo {volume} {09}},\ \bibinfo
  {pages} {033} (\bibinfo {year} {2017})},\ \Eprint
  {http://arxiv.org/abs/1611.05879} {arXiv:1611.05879 [hep-ph]} \BibitemShut
  {NoStop}%
\bibitem [{\citenamefont {Burdman}\ \emph {et~al.}(2015)\citenamefont
  {Burdman}, \citenamefont {Chacko}, \citenamefont {Harnik}, \citenamefont
  {de~Lima},\ and\ \citenamefont {Verhaaren}}]{Burdman:2014zta}%
  \BibitemOpen
  \bibfield  {author} {\bibinfo {author} {\bibfnamefont {G.}~\bibnamefont
  {Burdman}}, \bibinfo {author} {\bibfnamefont {Z.}~\bibnamefont {Chacko}},
  \bibinfo {author} {\bibfnamefont {R.}~\bibnamefont {Harnik}}, \bibinfo
  {author} {\bibfnamefont {L.}~\bibnamefont {de~Lima}}, \ and\ \bibinfo
  {author} {\bibfnamefont {C.~B.}\ \bibnamefont {Verhaaren}},\ }\href {\doibase
  10.1103/PhysRevD.91.055007} {\bibfield  {journal} {\bibinfo  {journal} {Phys.
  Rev.}\ }\textbf {\bibinfo {volume} {D91}},\ \bibinfo {pages} {055007}
  (\bibinfo {year} {2015})},\ \Eprint {http://arxiv.org/abs/1411.3310}
  {arXiv:1411.3310 [hep-ph]} \BibitemShut {NoStop}%
\bibitem [{\citenamefont {Foot}\ \emph {et~al.}(1991)\citenamefont {Foot},
  \citenamefont {Lew},\ and\ \citenamefont {Volkas}}]{Foot:1991bp}%
  \BibitemOpen
  \bibfield  {author} {\bibinfo {author} {\bibfnamefont {R.}~\bibnamefont
  {Foot}}, \bibinfo {author} {\bibfnamefont {H.}~\bibnamefont {Lew}}, \ and\
  \bibinfo {author} {\bibfnamefont {R.~R.}\ \bibnamefont {Volkas}},\ }\href
  {\doibase 10.1016/0370-2693(91)91013-L} {\bibfield  {journal} {\bibinfo
  {journal} {Phys. Lett.}\ }\textbf {\bibinfo {volume} {B272}},\ \bibinfo
  {pages} {67} (\bibinfo {year} {1991})}\BibitemShut {NoStop}%
\bibitem [{\citenamefont {Foot}\ \emph {et~al.}(1992)\citenamefont {Foot},
  \citenamefont {Lew},\ and\ \citenamefont {Volkas}}]{Foot:1991py}%
  \BibitemOpen
  \bibfield  {author} {\bibinfo {author} {\bibfnamefont {R.}~\bibnamefont
  {Foot}}, \bibinfo {author} {\bibfnamefont {H.}~\bibnamefont {Lew}}, \ and\
  \bibinfo {author} {\bibfnamefont {R.~R.}\ \bibnamefont {Volkas}},\ }\href
  {\doibase 10.1142/S0217732392004031} {\bibfield  {journal} {\bibinfo
  {journal} {Mod. Phys. Lett.}\ }\textbf {\bibinfo {volume} {A7}},\ \bibinfo
  {pages} {2567} (\bibinfo {year} {1992})}\BibitemShut {NoStop}%
\bibitem [{\citenamefont {Fujii}\ \emph {et~al.}(2015)\citenamefont {Fujii}
  \emph {et~al.}}]{Fujii:2015jha}%
  \BibitemOpen
  \bibfield  {author} {\bibinfo {author} {\bibfnamefont {K.}~\bibnamefont
  {Fujii}} \emph {et~al.},\ }\href@noop {} {\  (\bibinfo {year} {2015})},\
  \Eprint {http://arxiv.org/abs/1506.05992} {arXiv:1506.05992 [hep-ex]}
  \BibitemShut {NoStop}%
\bibitem [{\citenamefont {Abramowicz}\ \emph {et~al.}(2017)\citenamefont
  {Abramowicz} \emph {et~al.}}]{Abramowicz:2016zbo}%
  \BibitemOpen
  \bibfield  {author} {\bibinfo {author} {\bibfnamefont {H.}~\bibnamefont
  {Abramowicz}} \emph {et~al.},\ }\href {\doibase
  10.1140/epjc/s10052-017-4968-5} {\bibfield  {journal} {\bibinfo  {journal}
  {Eur. Phys. J.}\ }\textbf {\bibinfo {volume} {C77}},\ \bibinfo {pages} {475}
  (\bibinfo {year} {2017})},\ \Eprint {http://arxiv.org/abs/1608.07538}
  {arXiv:1608.07538 [hep-ex]} \BibitemShut {NoStop}%
\bibitem [{\citenamefont {Buttazzo}\ \emph {et~al.}(2015)\citenamefont
  {Buttazzo}, \citenamefont {Sala},\ and\ \citenamefont
  {Tesi}}]{Buttazzo:2015bka}%
  \BibitemOpen
  \bibfield  {author} {\bibinfo {author} {\bibfnamefont {D.}~\bibnamefont
  {Buttazzo}}, \bibinfo {author} {\bibfnamefont {F.}~\bibnamefont {Sala}}, \
  and\ \bibinfo {author} {\bibfnamefont {A.}~\bibnamefont {Tesi}},\ }\href
  {\doibase 10.1007/JHEP11(2015)158} {\bibfield  {journal} {\bibinfo  {journal}
  {JHEP}\ }\textbf {\bibinfo {volume} {11}},\ \bibinfo {pages} {158} (\bibinfo
  {year} {2015})},\ \Eprint {http://arxiv.org/abs/1505.05488} {arXiv:1505.05488
  [hep-ph]} \BibitemShut {NoStop}%
\bibitem [{\citenamefont {Ahmed}(2017)}]{Ahmed:2017psb}%
  \BibitemOpen
  \bibfield  {author} {\bibinfo {author} {\bibfnamefont {A.}~\bibnamefont
  {Ahmed}},\ }\href@noop {} {\  (\bibinfo {year} {2017})},\ \Eprint
  {http://arxiv.org/abs/1711.03107} {arXiv:1711.03107 [hep-ph]} \BibitemShut
  {NoStop}%
\bibitem [{\citenamefont {Behnke}\ \emph {et~al.}(2013)\citenamefont {Behnke},
  \citenamefont {Brau}, \citenamefont {Foster}, \citenamefont {Fuster},
  \citenamefont {Harrison}, \citenamefont {Paterson}, \citenamefont {Peskin},
  \citenamefont {Stanitzki}, \citenamefont {Walker},\ and\ \citenamefont
  {Yamamoto}}]{Behnke:2013xla}%
  \BibitemOpen
  \bibfield  {author} {\bibinfo {author} {\bibfnamefont {T.}~\bibnamefont
  {Behnke}}, \bibinfo {author} {\bibfnamefont {J.~E.}\ \bibnamefont {Brau}},
  \bibinfo {author} {\bibfnamefont {B.}~\bibnamefont {Foster}}, \bibinfo
  {author} {\bibfnamefont {J.}~\bibnamefont {Fuster}}, \bibinfo {author}
  {\bibfnamefont {M.}~\bibnamefont {Harrison}}, \bibinfo {author}
  {\bibfnamefont {J.~M.}\ \bibnamefont {Paterson}}, \bibinfo {author}
  {\bibfnamefont {M.}~\bibnamefont {Peskin}}, \bibinfo {author} {\bibfnamefont
  {M.}~\bibnamefont {Stanitzki}}, \bibinfo {author} {\bibfnamefont
  {N.}~\bibnamefont {Walker}}, \ and\ \bibinfo {author} {\bibfnamefont
  {H.}~\bibnamefont {Yamamoto}},\ }\href@noop {} {\  (\bibinfo {year}
  {2013})},\ \Eprint {http://arxiv.org/abs/1306.6327} {arXiv:1306.6327
  [physics.acc-ph]} \BibitemShut {NoStop}%
\bibitem [{\citenamefont {Boland}\ \emph {et~al.}(2016)\citenamefont {Boland}
  \emph {et~al.}}]{CLIC:2016zwp}%
  \BibitemOpen
  \bibfield  {author} {\bibinfo {author} {\bibfnamefont {M.~J.}\ \bibnamefont
  {Boland}} \emph {et~al.} (\bibinfo {collaboration} {CLICdp, CLIC}),\ }\href
  {\doibase 10.5170/CERN-2016-004} {\  (\bibinfo {year} {2016}),\
  10.5170/CERN-2016-004},\ \Eprint {http://arxiv.org/abs/1608.07537}
  {arXiv:1608.07537 [physics.acc-ph]} \BibitemShut {NoStop}%
\bibitem [{\citenamefont {Aad}\ \emph {et~al.}(2016)\citenamefont {Aad} \emph
  {et~al.}}]{Khachatryan:2016vau}%
  \BibitemOpen
  \bibfield  {author} {\bibinfo {author} {\bibfnamefont {G.}~\bibnamefont
  {Aad}} \emph {et~al.} (\bibinfo {collaboration} {ATLAS, CMS}),\ }\href
  {\doibase 10.1007/JHEP08(2016)045} {\bibfield  {journal} {\bibinfo  {journal}
  {JHEP}\ }\textbf {\bibinfo {volume} {08}},\ \bibinfo {pages} {045} (\bibinfo
  {year} {2016})},\ \Eprint {http://arxiv.org/abs/1606.02266} {arXiv:1606.02266
  [hep-ex]} \BibitemShut {NoStop}%
\bibitem [{\citenamefont {Dawson}\ \emph {et~al.}(2013)\citenamefont {Dawson}
  \emph {et~al.}}]{Dawson:2013bba}%
  \BibitemOpen
  \bibfield  {author} {\bibinfo {author} {\bibfnamefont {S.}~\bibnamefont
  {Dawson}} \emph {et~al.},\ }in\ \href
  {https://inspirehep.net/record/1262795/files/arXiv:1310.8361.pdf} {\emph
  {\bibinfo {booktitle} {{Proceedings, 2013 Community Summer Study on the
  Future of U.S. Particle Physics: Snowmass on the Mississippi (CSS2013):
  Minneapolis, MN, USA, July 29-August 6, 2013}}}}\ (\bibinfo {year} {2013})\
  \Eprint {http://arxiv.org/abs/1310.8361} {arXiv:1310.8361 [hep-ex]}
  \BibitemShut {NoStop}%
\bibitem [{\citenamefont {{ATLAS
  Collaboration}}(2013)}]{ATL-PHYS-PUB-2013-016}%
  \BibitemOpen
  \bibfield  {author} {\bibinfo {author} {\bibnamefont {{ATLAS Collaboration}}}
  (\bibinfo {collaboration} {ATLAS}),\ }\href
  {http://cds.cern.ch/record/1611190} {\  (\bibinfo {year} {2013})},\ \Eprint
  {http://arxiv.org/abs/ATL-PHYS-PUB-2013-016} {ATL-PHYS-PUB-2013-016}
  \BibitemShut {NoStop}%
\bibitem [{\citenamefont {{CMS Collaboration}}(2013)}]{CMS:2013dga}%
  \BibitemOpen
  \bibfield  {author} {\bibinfo {author} {\bibnamefont {{CMS Collaboration}}}
  (\bibinfo {collaboration} {CMS}),\ }\href@noop {} {\  (\bibinfo {year}
  {2013})},\ \Eprint {http://arxiv.org/abs/CMS-PAS-FTR-13-024}
  {CMS-PAS-FTR-13-024} \BibitemShut {NoStop}%
\bibitem [{\citenamefont {Holzner}(2014)}]{Holzner:2014qqs}%
  \BibitemOpen
  \bibfield  {author} {\bibinfo {author} {\bibfnamefont {A.}~\bibnamefont
  {Holzner}} (\bibinfo {collaboration} {ATLAS, CMS}),\ }\href@noop {} {\
  (\bibinfo {year} {2014})},\ \Eprint {http://arxiv.org/abs/1411.0322}
  {arXiv:1411.0322 [hep-ex]} \BibitemShut {NoStop}%
\bibitem [{\citenamefont {de~Florian}\ \emph {et~al.}(2016)\citenamefont
  {de~Florian} \emph {et~al.}}]{deFlorian:2016spz}%
  \BibitemOpen
  \bibfield  {author} {\bibinfo {author} {\bibfnamefont {D.}~\bibnamefont
  {de~Florian}} \emph {et~al.} (\bibinfo {collaboration} {LHC Higgs Cross
  Section Working Group}),\ }\href@noop {} {\  (\bibinfo {year} {2016})},\
  \Eprint {http://arxiv.org/abs/1610.07922} {arXiv:1610.07922 [hep-ph]}
  \BibitemShut {NoStop}%
\bibitem [{\citenamefont {L{HC Higgs Cross Section Working
  Group}}(2017)}]{HWG}%
  \BibitemOpen
  \bibfield  {author} {\bibinfo {author} {\bibnamefont {L{HC Higgs Cross
  Section Working Group}}},\ }\href@noop {} {}\bibinfo {howpublished}
  {\href{https://twiki.cern.ch/twiki/bin/view/LHCPhysics/LHCHXSWG}
  {https://twiki.cern.ch/twiki/bin/view/LHCPhysics/LHCHXSWG}} (\bibinfo {year}
  {2017})\BibitemShut {NoStop}%
\bibitem [{\citenamefont {H{iggs cross sections for HL-LHC and
  HE-LHC}}(2016)}]{PDFratio}%
  \BibitemOpen
  \bibfield  {author} {\bibinfo {author} {\bibnamefont {H{iggs cross sections
  for HL-LHC and HE-LHC}}},\ }\href@noop {} {}\bibinfo {howpublished}
  {\href{https://twiki.cern.ch/twiki/bin/view/LHCPhysics/HiggsEuropeanStrategy}
  {https://twiki.cern.ch/twiki/bin/view/LHCPhysics/HiggsEuropeanStrategy}}
  (\bibinfo {year} {2016})\BibitemShut {NoStop}%
\bibitem [{\citenamefont {Mangano}\ \emph {et~al.}(2017)\citenamefont {Mangano}
  \emph {et~al.}}]{Mangano:2016jyj}%
  \BibitemOpen
  \bibfield  {author} {\bibinfo {author} {\bibfnamefont {M.~L.}\ \bibnamefont
  {Mangano}} \emph {et~al.},\ }\href@noop {} {\bibfield  {journal} {\bibinfo
  {journal} {CERN Yellow Report}\ ,\ \bibinfo {pages} {1}} (\bibinfo {year}
  {2017})},\ \Eprint {http://arxiv.org/abs/1607.01831} {arXiv:1607.01831
  [hep-ph]} \BibitemShut {NoStop}%
\bibitem [{\citenamefont {Alwall}\ \emph {et~al.}(2014)\citenamefont {Alwall},
  \citenamefont {Frederix}, \citenamefont {Frixione}, \citenamefont {Hirschi},
  \citenamefont {Maltoni}, \citenamefont {Mattelaer}, \citenamefont {Shao},
  \citenamefont {Stelzer}, \citenamefont {Torrielli},\ and\ \citenamefont
  {Zaro}}]{Alwall:2014hca}%
  \BibitemOpen
  \bibfield  {author} {\bibinfo {author} {\bibfnamefont {J.}~\bibnamefont
  {Alwall}}, \bibinfo {author} {\bibfnamefont {R.}~\bibnamefont {Frederix}},
  \bibinfo {author} {\bibfnamefont {S.}~\bibnamefont {Frixione}}, \bibinfo
  {author} {\bibfnamefont {V.}~\bibnamefont {Hirschi}}, \bibinfo {author}
  {\bibfnamefont {F.}~\bibnamefont {Maltoni}}, \bibinfo {author} {\bibfnamefont
  {O.}~\bibnamefont {Mattelaer}}, \bibinfo {author} {\bibfnamefont {H.~S.}\
  \bibnamefont {Shao}}, \bibinfo {author} {\bibfnamefont {T.}~\bibnamefont
  {Stelzer}}, \bibinfo {author} {\bibfnamefont {P.}~\bibnamefont {Torrielli}},
  \ and\ \bibinfo {author} {\bibfnamefont {M.}~\bibnamefont {Zaro}},\ }\href
  {\doibase 10.1007/JHEP07(2014)079} {\bibfield  {journal} {\bibinfo  {journal}
  {JHEP}\ }\textbf {\bibinfo {volume} {07}},\ \bibinfo {pages} {079} (\bibinfo
  {year} {2014})},\ \Eprint {http://arxiv.org/abs/1405.0301} {arXiv:1405.0301
  [hep-ph]} \BibitemShut {NoStop}%
\bibitem [{\citenamefont {Sj{\"o}strand}\ \emph {et~al.}(2015)\citenamefont
  {Sj{\"o}strand}, \citenamefont {Ask}, \citenamefont {Christiansen},
  \citenamefont {Corke}, \citenamefont {Desai}, \citenamefont {Ilten},
  \citenamefont {Mrenna}, \citenamefont {Prestel}, \citenamefont {Rasmussen},\
  and\ \citenamefont {Skands}}]{Sjostrand:2014zea}%
  \BibitemOpen
  \bibfield  {author} {\bibinfo {author} {\bibfnamefont {T.}~\bibnamefont
  {Sj{\"o}strand}}, \bibinfo {author} {\bibfnamefont {S.}~\bibnamefont {Ask}},
  \bibinfo {author} {\bibfnamefont {J.~R.}\ \bibnamefont {Christiansen}},
  \bibinfo {author} {\bibfnamefont {R.}~\bibnamefont {Corke}}, \bibinfo
  {author} {\bibfnamefont {N.}~\bibnamefont {Desai}}, \bibinfo {author}
  {\bibfnamefont {P.}~\bibnamefont {Ilten}}, \bibinfo {author} {\bibfnamefont
  {S.}~\bibnamefont {Mrenna}}, \bibinfo {author} {\bibfnamefont
  {S.}~\bibnamefont {Prestel}}, \bibinfo {author} {\bibfnamefont {C.~O.}\
  \bibnamefont {Rasmussen}}, \ and\ \bibinfo {author} {\bibfnamefont {P.~Z.}\
  \bibnamefont {Skands}},\ }\href {\doibase 10.1016/j.cpc.2015.01.024}
  {\bibfield  {journal} {\bibinfo  {journal} {Comput. Phys. Commun.}\ }\textbf
  {\bibinfo {volume} {191}},\ \bibinfo {pages} {159} (\bibinfo {year}
  {2015})},\ \Eprint {http://arxiv.org/abs/1410.3012} {arXiv:1410.3012
  [hep-ph]} \BibitemShut {NoStop}%
\bibitem [{\citenamefont {de~Favereau}\ \emph {et~al.}(2014)\citenamefont
  {de~Favereau}, \citenamefont {Delaere}, \citenamefont {Demin}, \citenamefont
  {Giammanco}, \citenamefont {Lema\^{i}tre}, \citenamefont {Mertens},\ and\
  \citenamefont {Selvaggi}}]{deFavereau:2013fsa}%
  \BibitemOpen
  \bibfield  {author} {\bibinfo {author} {\bibfnamefont {J.}~\bibnamefont
  {de~Favereau}}, \bibinfo {author} {\bibfnamefont {C.}~\bibnamefont
  {Delaere}}, \bibinfo {author} {\bibfnamefont {P.}~\bibnamefont {Demin}},
  \bibinfo {author} {\bibfnamefont {A.}~\bibnamefont {Giammanco}}, \bibinfo
  {author} {\bibfnamefont {V.}~\bibnamefont {Lema\^{i}tre}}, \bibinfo {author}
  {\bibfnamefont {A.}~\bibnamefont {Mertens}}, \ and\ \bibinfo {author}
  {\bibfnamefont {M.}~\bibnamefont {Selvaggi}} (\bibinfo {collaboration}
  {DELPHES 3}),\ }\href {\doibase 10.1007/JHEP02(2014)057} {\bibfield
  {journal} {\bibinfo  {journal} {JHEP}\ }\textbf {\bibinfo {volume} {02}},\
  \bibinfo {pages} {057} (\bibinfo {year} {2014})},\ \Eprint
  {http://arxiv.org/abs/1307.6346} {arXiv:1307.6346 [hep-ex]} \BibitemShut
  {NoStop}%
\bibitem [{\citenamefont {Cacciari}\ \emph {et~al.}(2008)\citenamefont
  {Cacciari}, \citenamefont {Salam},\ and\ \citenamefont
  {Soyez}}]{Cacciari:2008gp}%
  \BibitemOpen
  \bibfield  {author} {\bibinfo {author} {\bibfnamefont {M.}~\bibnamefont
  {Cacciari}}, \bibinfo {author} {\bibfnamefont {G.~P.}\ \bibnamefont {Salam}},
  \ and\ \bibinfo {author} {\bibfnamefont {G.}~\bibnamefont {Soyez}},\ }\href
  {\doibase 10.1088/1126-6708/2008/04/063} {\bibfield  {journal} {\bibinfo
  {journal} {JHEP}\ }\textbf {\bibinfo {volume} {04}},\ \bibinfo {pages} {063}
  (\bibinfo {year} {2008})},\ \Eprint {http://arxiv.org/abs/0802.1189}
  {arXiv:0802.1189 [hep-ph]} \BibitemShut {NoStop}%
\bibitem [{\citenamefont {Cacciari}\ \emph {et~al.}(2012)\citenamefont
  {Cacciari}, \citenamefont {Salam},\ and\ \citenamefont
  {Soyez}}]{Cacciari:2011ma}%
  \BibitemOpen
  \bibfield  {author} {\bibinfo {author} {\bibfnamefont {M.}~\bibnamefont
  {Cacciari}}, \bibinfo {author} {\bibfnamefont {G.~P.}\ \bibnamefont {Salam}},
  \ and\ \bibinfo {author} {\bibfnamefont {G.}~\bibnamefont {Soyez}},\ }\href
  {\doibase 10.1140/epjc/s10052-012-1896-2} {\bibfield  {journal} {\bibinfo
  {journal} {Eur. Phys. J.}\ }\textbf {\bibinfo {volume} {C72}},\ \bibinfo
  {pages} {1896} (\bibinfo {year} {2012})},\ \Eprint
  {http://arxiv.org/abs/1111.6097} {arXiv:1111.6097 [hep-ph]} \BibitemShut
  {NoStop}%
\bibitem [{\citenamefont {Abramowicz}\ \emph {et~al.}(2013)\citenamefont
  {Abramowicz} \emph {et~al.}}]{Behnke:2013lya}%
  \BibitemOpen
  \bibfield  {author} {\bibinfo {author} {\bibfnamefont {H.}~\bibnamefont
  {Abramowicz}} \emph {et~al.},\ }\href@noop {} {\  (\bibinfo {year} {2013})},\
  \Eprint {http://arxiv.org/abs/1306.6329} {arXiv:1306.6329 [physics.ins-det]}
  \BibitemShut {NoStop}%
\bibitem [{\citenamefont {Djouadi}(2008)}]{Djouadi:2005gi}%
  \BibitemOpen
  \bibfield  {author} {\bibinfo {author} {\bibfnamefont {A.}~\bibnamefont
  {Djouadi}},\ }\href {\doibase 10.1016/j.physrep.2007.10.004} {\bibfield
  {journal} {\bibinfo  {journal} {Phys. Rept.}\ }\textbf {\bibinfo {volume}
  {457}},\ \bibinfo {pages} {1} (\bibinfo {year} {2008})},\ \Eprint
  {http://arxiv.org/abs/hep-ph/0503172} {arXiv:hep-ph/0503172 [hep-ph]}
  \BibitemShut {NoStop}%
\end{thebibliography}%

\end{document}